\documentclass[]{aastex631}

\usepackage{amsmath}
\usepackage{amssymb}
\begin{document}

\title{\uppercase{Monte Carlo Simulations of Polarized Radiative Transfer\\ in Neutron Star Atmospheres}}

\author[0000-0001-9268-5577]{Hoa Dinh Thi}
\affiliation{Department of Physics and Astronomy—MS 108, Rice University, 6100 Main St., Houston, TX 77251-1892, USA}

%\collaboration{20}{(AAS Journals Data Editors)}

%
%         ********ESSENTIAL FONT DEFINITIONS********
%
\font\fiverm=cmr5       \font\fivebf=cmbx5      \font\sevenrm=cmr7     
\font\sevenbf=cmbx7     \font\eightrm=cmr8      \font\eighti=cmmi8     
\font\eightsy=cmsy8     \font\eightbf=cmbx8     \font\eighttt=cmtt8    
\font\eightit=cmti8     \font\eightsl=cmsl8     \font\sixrm=cmr6       
\font\sixi=cmmi6        \font\sixsy=cmsy6       \font\sixbf=cmbx6      
\def\dover#1#2{\hbox{${{\displaystyle#1 \vphantom{(} }\over{
   \displaystyle #2 \vphantom{(} }}$}}

\def\fsc{\alpha_{\hbox{\sevenrm f}}}                                
\def\sigt{\sigma_{\hbox{\fiverm T}} }                                
\def\taut{\tau_{\hbox{\fiverm T}}}                                
\def\lambar{\lambda\llap {--}}
\def\omegaB{\omega_{\hbox{\sixrm B}}}
\def\SigmaB{\Sigma_{\hbox{\sixrm B}}}
\def\DeltaB{\Delta_{\hbox{\sixrm B}}}
\def\wcyc{\omega_{\hbox{\fiverm B}}}
\def\thetaMSC{\theta_{\hbox{\sixrm MSC}}}

\def\hruleseparator{\vskip 8pt \textcolor{blue}{\centerline{\vbox{\hrule height 2.0pt width 100pt}}} \vskip 10pt}

\def\teq#1{$\, #1\,$}                         % text equation
\def\edithere#1{\textcolor{red}{#1}}  % flag for editing (red)
\def\actionitem#1{\textcolor{blue}{#1}}  % flag for editing (blue)
\def\commenthere#1{\textcolor{green}{#1}}  % flag for editing (green)

\author[0000-0003-4433-1365]{Matthew G. Baring}
\affiliation{Department of Physics and Astronomy—MS 108, Rice University, 6100 Main St., Houston, TX 77251-1892, USA}

\author[0000-0002-9705-7948]{Kun Hu}
\affiliation{Physics Department, McDonnell Center for the Space Sciences, and Center for Quantum Leaps, Washington University in St. Louis, St. Louis, MO 63130, USA}

\author[0000-0001-6119-859X]{Alice K. Harding}
\affiliation{Theoretical Division, Los Alamos National Laboratory, Los Alamos, NM 58545, USA}

\author[0000-0003-0503-0914]{Joseph A. Barchas}
\affiliation{Natural Sciences, Southwest Campus, Houston Community College, 5601 W. Loop S., Houston, TX 77081, USA}

%\author[0000-0002-7991-028X]{George A. Younes}
%\affiliation{Astrophysics Science Division, NASA Goddard Space Flight Center, Greenbelt, MD 20771, USA}
%\affiliation{Department of Astronomy, University of
	%Maryland College Park, College Park,
	%MD 20742, USA}

%\author[0000-0002-0254-5915]{Rachael Stewart}
%\affiliation{Department of Physics, George Washington University 1918 F Street, NW Washington, DC 20052 USA}


\begin{abstract}
Soft X-ray emission from neutron stars affords powerful diagnostic tools for uncovering their surface and interior properties, as well as their geometric configurations.  In the atmospheres of neutron stars, the presence of magnetic fields alters the photon-electron scattering cross sections, resulting in non-trivial angular dependence of intensity and polarization of the emergent signals. This paper presents recent developments of our Monte Carlo simulation, {\sl MAGTHOMSCATT}, which tracks the complex electric field vector for each photon during its transport.  Our analysis encompasses the anisotropy and polarization characteristics of X-ray emission for field strengths ranging from non-magnetic to extremely magnetized regimes that are germane to magnetars. In the very low field domain, we reproduced the numerical solution to the radiative transfer equation for non-magnetic Thomson scattering, and provided analytical fits for the angular dependence of the intensity and the polarization degree. These fits can be useful for studies of millisecond pulsars and magnetic white dwarfs.  By implementing a refined injection protocol, we show that, in the magnetar regime, the simulated intensity and polarization pulse profiles of emission from extended surface regions becomes invariant with respect to the ratio of photon ($\omega$) and electron cyclotron ($\wcyc$) frequencies once $\omega / \wcyc \lesssim 0.01$. This circumvents the need for simulations pertinent to really high magnetic field strengths, which are inherently slower. Our approach will be employed elsewhere to model observational data to constrain neutron star geometric parameters and properties of emitting hot spots on their surfaces.

\end{abstract}



%% cross-references to sections, equations, tables, and figure*s.

\section{Introduction}
 \label{sec:intro}
The modeling of surface X-ray emission in neutron stars, which is strongly polarized by their intense magnetic fields $\boldsymbol{B}$, has received extensive treatment over the years.  A common feature of most studies is that they use magnetic Thomson scattering and free-free opacity to mediate the photon transport and also help support the atmospheres hydrostatically.  Older investigations focused on traditional neutron stars with surface polar fields \teq{B_p < 10^{13}}Gauss, and mostly presumed fully-ionized hydrogen/helium atmospheres at temperatures \teq{T\sim 10^6}K \citep{Shibanov-1992-AandA,Pavlov-1994-AandA,Zavlin-1996-AandA,Zane-2000-ApJ}. Yet, partially-ionized atmosphere models have also emerged \citep[e.g.,][]{HLPC-2003-ApJ,Potekhin-2004-ApJ,Suleimanov-2009-AandA}. Higher magnetizations are inferred for the active subset of pulsars now known as magnetars, with spin-down field estimates typically in the \teq{B_p \sim 10^{14} - 2 \times 10^{15}}Gauss range.  Addressing magnetar domains required a new type of atmosphere models \citep[e.g.,][]{Ho-Lai-2001-MNRAS,Ozel-2001-ApJ,Ozel-2003-ApJ,Adelsberg-2006-MNRAS} that included detailed treatment of the polarization of the vacuum by the field \citep{Tsai-1975-PhRvD}.  This quantum electrodynamical (QED) influence renders the vacuum birefringent, so that radiative transfer is sensitive to the polarization state (i.e., radiation eigenmode), just as it is for a plasma, although the preferred mode differs between plasma-dominated and vacuum-dominated regimes. This birefringence physics also dictates how the polarized radiation signals propagate to the distant observer \citep[e.g.,][]{Heyl-2000-MNRAS,Lai-Ho-2003PhRvL,Fernandez-2011-ApJ, Taverna-2015-MNRAS}: electric field vectors of light rotate in the varying magnetic field.

When emission is not uniform across a neutron star surface, stellar rotation leads to soft X-ray intensity pulsation, as is observed for different classes of pulsars.  For a given neutron star, this enables powerful diagnostics on the rotator inclination (\teq{\alpha}) and observer viewing (\teq{\zeta}) angles relative to the rotation axis \teq{\boldsymbol{\Omega}}.  The case of the compact central object (CCO) pulsar RX J0822-4300 with its weak magnetization \citep[\teq{B_p = 5.7 \times 10^{10}}Gauss; see][]{Gotthelf-2013-ApJ} was studied by \citep{Gotthelf2010}, and later by \citep{Alford-2022-ApJ}, with the conclusion that two hot spots of different temperatures radiated the observed X rays from approximately antipodal locales. For magnetars, which are generally hotter and therefore brighter than conventional pulsars, several groups have explored geometrical diagnostics using pulse profile information.  For instance, \cite{Albano-2010-ApJ,Bernardini-2011-MNRAS} fit the pulse profiles of XTE~J1810-197 during quiescence while \cite{Albano-2010-ApJ} did the same for CXOU~J164710.2-455216 (finding \teq{\alpha\sim 80^{\circ}} and \teq{\zeta\sim 25^{\circ}}), employing blackbody emission from the whole surface with a prescribed colatitudinal temperature distribution. Both works included gravitational light bending.  With a similar construct, \cite{Guillot-2015-MNRAS} found that a very small polar region with a temperature a factor of 6 larger than that of the equatorial zone is required to fit the single-peaked soft X-ray profile for SGR~0418+572.  Similarly, \cite{Younes-2020-ApJ} fit the double-peaked pulse profile of a 2.2 hour flare detected from 1RXS J1708-4009 with \teq{\alpha\sim 60^{\circ}} and \teq{\zeta\sim 60^{\circ}}, concluding that antipodal spots are required to match the soft X-ray light curve during the flare episode. All of these studies omitted detailed treatments of local radiation anisotropy, which varies in atmospheres as the direction and magnitude of $\boldsymbol{B}$ change across the stellar surface.

Adding modulated Stokes' parameter polarization profiles for rotating neutron stars into the mix offers the prospect of more tightly constraining stellar parameters. Presently this is only possible for the hotter and brighter magnetars, for five of which the acquisition of X-ray polarization signals has been provided by the Imaging X-ray Polarimetry Explorer ({\it IXPE}) in its sensitive 2-8 keV band. For 1RXS J1708-4009 \citep{Zane-2023-ApJ}, the pulse profile is single-peaked at \teq{<4}keV energies, and the linear polarization degree (PD) is anti-correlated with the intensity. Interestingly, its polarization angle (PA) was essentially the same at all energies. This ``fixed'' PA strongly contrasts the wide-ranging {\it IXPE}~PA values seen in 4U 0142+61 between $<4\;$keV energies, where the PD is 14\%, and \teq{>5.5}keV energies, where the PD is 42\%  \citep{Taverna-2022-Sci}. The \teq{90^{\circ}} PA dichotomy demarcated by these energy bands is perhaps most naturally interpreted \citep{Taverna-2022-Sci} as the \teq{>5.5}keV signal emanating from an inner magnetospheric coronal region (as opposed to the surface) where resonant cyclotron scattering of surface photons occurs. SGR 1806-20 provided a further puzzle, with significant PD levels of 40\% only in the 4-5 keV window, and less than 5\% below 4 keV, leading to the suggestion \citep{Turolla-2023-ApJ} that the thermal emission may originate from a condensed part of the stellar surface.  Low polarization degrees were predicted by condensed magnetar surface models \citep{Taverna-2020-MNRAS}, yet can also result from polarization mode switching at the vacuum resonance in ionized atmospheres \citep{Lai-2023-PNAS}.

To determine the expectations for pulse profiles from atmospheres of rotating neutron stars, it is necessary to treat a broad range of surface locales that possess varied magnetic field strengths and directions.   Monte Carlo simulations \citep[e.g.,][]{Bulik-1997-MNRAS,Niemiec-2006-ApJ,Barchas-2021-mnras} afford a versatility for treatment of atmospheric radiative transport spanning a variety of magnetic colatitudes.  They complement and contrast semi-analytic/numerical protocols \citep[e.g.,][]{Ho-Lai-2001-MNRAS,Ozel-2001-ApJ} for the solution of the radiative transfer equation that are best suited to the magnetic pole, where the field coincides with the local zenith direction.  To pursue this objective, our group developed a Monte Carlo simulation named {\sl MAGTHOMSCATT}  that tracks polarized photon propagation and scattering in strong fields threading atmospheric slabs, encapsulating a broad range of magnetic field domains.  The baseline technical details are described in \cite{Barchas-2021-mnras,Hu-2022-ApJ}, and sample results that include the influence of general relativity on propagation of light to infinity can be found in \cite{Hu-2022-ApJ}.  The code simulates polarized magnetic Thomson scattering due to electrons in arbitrary (usually, but not necessarily uniform) magnetic field configurations for an optically thick, fully ionized gas; its structure in optical depth units enables it to address a wide variety of radial density gradients.  {\sl MAGTHOMSCATT} has been validated extensively \citep{Barchas-2021-mnras}, reproducing anisotropic angular distributions reported by \cite{Whitney-1991-ApJS} for fields $\boldsymbol{B}$ aligned with the zenith direction, and non-magnetic transport characteristics detailed in the papers of \cite{Sunyaev-1980-AandA,Sunyaev-1985-AandA}. The atmospheric transport simulation yields anisotropies that depend on the local field direction, eliciting profound differences in anisotropy and polarization signatures in surface regions spanning the magnetic pole to the equator.

In this paper, a deeper exploration of the intricacies of {\sl MAGTHOMSCATT} and its broad range of precision modeling is expounded. It will become evident that well-designed, modest algorithmic upgrades enable the simulation to be applied to neutron stars and white dwarfs spanning 6-8 orders of magnitude in magnetic field strength.  After a brief summary of its structure, in Section 2 the distributions for the coupling between scattering number and vertical altitude will be explored, contrasting populations emerging from the upper surface of the atmospheric slab with those escaping through the bottom boundary. Section~3 details the anisotropy and polarization characteristics of radiation emergent from the upper surface into the magnetosphere, spanning highly-magnetic and essentially non-magnetic domains. As the code is inherently slower for runs near the magnetic pole when the field strength is really high, Section~4 outlines a revised injection protocol that enhances simulation efficiency, typically leading to run time reductions by factors of 2-8.  In addition, it is shown that pulse profile information in both intensity and polarization degree resulting from extended polar caps saturates once the field becomes sufficiently strong; this informs run protocols for accurately capturing expected pulsation signals for magnetars.

\section{Polarized Radiative Transfer in the Magnetic Thomson Domain}
\label{sec:radiative-transfer}

As a prelude to the detailed results in Section~\ref{sec:emergent} on the  intensity and polarization characteristics emergent from the simulation of neutron star localized surface regions, here the exposition summarizes the scattering physics cross section and the main elements of the radiative transfer simulation.  It then focuses on the properties of photon diffusion within the atmospheric slabs, which helps determine protocols for improving the run efficiency of the Monte Carlo simulation.

\subsection{Magnetic Thomson Cross Sections}
\label{subsec:magThomson}

The intense magnetic field in magnetars and other isolated neutron stars plays an essential role in the anisotropy and polarization properties of X-ray emission from their surfaces. The photon frequency ratios pertinent to magnetar $< 5$~keV surface signals correspond to \teq{\omega /\wcyc \ll 1 }, where $\omega$ and \teq{\wcyc} are respectively the photon and electron cyclotron frequencies.   Magnetic Thomson scattering is classical in this domain, with QED modifications only becoming important for photon energies above around 30~keV when electron recoil and Klein-Nishina influences become significant. The cross section is resonant at the electron cyclotron frequency \teq{\wcyc = eB/(m_ec)}, with $e$, $m_e$, and $c$ respectively being the electron charge and mass, and the speed of light, and $B = | \boldsymbol{B}|$ is the magnetic field strength.
Equation~(B9) in \cite{Barchas-2021-mnras} exhibits the polarization-dependent differential cross section that is relevant to our work. After integrating over the final azimuthal angle, the differential cross sections for all the polarization modes are
\begin{eqnarray}
   \dover{d\sigma_{\perp \rightarrow \perp}}{d\mu_f} \; =\; \pi r_0^2\, \SigmaB (\omega )
   \quad & , & \quad
   \dover{d\sigma_{\perp \rightarrow \parallel}}{d\mu_f} 
   \; =\; \pi r_0^2\, \SigmaB (\omega ) \mu_f^2 \nonumber \\[-5.5pt]
 \label{eq:dsig_magThom} \\[-5.5pt]
   \dover{d\sigma_{\parallel \rightarrow \perp}}{d\mu_f} \; =\; \pi r_0^2\, \SigmaB (\omega ) \mu_i^2
   \quad & , & \quad
   \dover{d\sigma_{\parallel \rightarrow \parallel}}{d\mu_f} \; =\; 
   \pi r_0^2 \,\Bigl[ 2 (1-\mu_i^2) (1-\mu_f^2) + \SigmaB (\omega )\, \mu_i^2\mu_f^2 \Bigr] \quad . \nonumber
\end{eqnarray}
These forms are expressed compactly using a frequency-dependent function that encapsulates the resonant character:
\begin{equation}
    \SigmaB (\omega ) \; =\; \dover{\omega^2(\omega^2+\wcyc^2)}{ (\omega^2 - \wcyc^2)^2}
    \; =\; \dover{1}{2} \biggl\{ \dover{\omega^2}{(\omega - \wcyc)^2}  
         + \dover{\omega^2}{(\omega + \wcyc)^2} \biggr\}
	\; \approx\; \dover{\omega^2}{\wcyc^2} \quad \hbox{for}\quad \omega \; \ll\; \wcyc \quad .
 \label{eq:SigmaB_def}
\end{equation}
For these cross sections, $\parallel$ ($\perp$) corresponds to the polarization state with the electric field vector being in (perpendicular to) the plane containing the propagation vector and magnetic field vector.  Also, $r_0 = e^2 /(m_ec^2) $ is the classical electron radius, and $\mu_f = \cos \theta_{f}$ and $\mu_i = \cos \theta_{i}$ are, respectively, the cosines of the angle between the scattered and incident unit wave vectors $\hat{\boldsymbol{k}}_f$ and $\hat{\boldsymbol{k}}_i$ and the magnetic field direction $\boldsymbol{B}$; see also Figure B1 in \citet{Barchas-2021-mnras}. From Equation~(\ref{eq:dsig_magThom}), it is evident that for $\omega /\wcyc \ll 1$, the cross section is generally dominated by the $\parallel \rightarrow \parallel$ contribution.  An exception to this arises for $\theta_{i} \lesssim 2\omega /\wcyc$, for which the differential cross sections for all the linear polarization modes are of similar magnitude.  Therefore, one can expect that the emerging radiation is beamed along the magnetic field direction within an angle of $\sim 2\thetaMSC$, where $\thetaMSC = \omega / \wcyc$ is the opening angle of the so-called magnetic scattering cone (MSC), a term coined by \citet{Baring-ApJ-2025}.

\subsection{Radiative Transfer Simulations: Summary of Key Elements} % at High Opacity} 
\label{subsec:magthomscatt}
The outer surface layer of a neutron star can be represented as a collection of thin atmospheric slabs, each corresponding to a magnetic field orientation characterized by $\theta_{\rm B} = \arccos\left(\hat{\boldsymbol{n}} \cdot \hat{\boldsymbol{B}}\right)$, with $\hat{\boldsymbol{n}}$ being the unit normal vector of the slab and $\hat{\boldsymbol{B}} =\boldsymbol{B} / |\boldsymbol{B} | $  the unit magnetic field vector. To model the scattering transport of polarized radiation through these localized atmospheric slabs, we employ the {\sl MAGTHOMSCATT} Monte Carlo simulation, which was initially developed by \citet{Barchas-2021-mnras} and further enhanced by \citet{Hu-2022-ApJ}. This simulation incorporates a complex electric field vector formalism, allowing for tracking the electric field vector $\boldsymbol{\mathcal{E}}$ of each photon, hence complete polarization characteristics (linearity, circularity, and ellipticity), throughout its scattering process in the slab and propagation through the magnetosphere. This approach differs from previous works that relied on tracking Stokes parameter information \citep[see, e.g.,][]{Whitney-1991-ApJS} or solving the radiative transfer equation for two normal modes \citep[see, e.g.,][]{Ho-Lai-2001-MNRAS,Ozel-2001-ApJ}. Therefore, it offers a more comprehensive understanding of neutron star atmospheric emission; see  \citet{Barchas-2021-mnras} and \citet{Hu-2022-ApJ} for details. Below, we outline the key points of the simulation.

In the simulation, photons are injected at the bottom of a localized atmospheric slab (that is uniform in optical depth) at a magnetic colatitude corresponding to $\theta_{\rm B}$. These photons then undergo magnetic Thomson scatterings, and only those that exit from the top of the surface layer are recorded. Upon successfully escaping from the neutron star surface, they propagate through a general relativistic magnetosphere, eventually reaching an observer at infinity; see Figure 1 of \citet{Hu-2022-ApJ} for an illustration of the geometry. We will present the details of the magnetospheric propagation and the modeling of observed intensity and polarization pulse profiles in 
an upcoming work, Dinh Thi et al. (in prep). In this paper, we will primarily focus on analyzing the  intensity and polarization characteristics of emission emergent on the neutron star surface, with the {\sl MAGTHOMSCATT} simulations performed in the local inertial frame germane to a Schwarzschild metric.

While one could opt to inject isotropic and unpolarized (IU) photons at the base of the atmospheric slab, this simple approach could require substantial computational time to simulate scatterings deep inside the atmosphere, particularly in the sub-cyclotronic domain. To overcome this drawback of the IU protocol, \citet{Barchas-2021-mnras} devised the so-called anisotropic and polarized (AP) injection protocol. Within this injection method, photons of frequency $\omega$ are initially generated according to the following angular distribution:
\begin{equation}
	I(\mu_0) \; =\; \frac{A_{\omega}(\mu_0)}{\Lambda_{\omega} \sigma(\omega, \mu_0)}
    \quad ,\quad
    \Lambda_{\omega} \; =\; 
    \int_{-1}^{1}\frac{A_{\omega}(\mu_0)}{\sigma(\omega, \mu_0)} d\mu_0 \quad ,
	\label{eq:I_mui}
\end{equation}
where $\sigma(\omega, \mu_0)$ is the total magnetic Thomson scattering cross section that includes polarization information; it is given by Equation~(13) in \citet{Barchas-2021-mnras}.  Therein, $\mu_0 = \hat{\boldsymbol{k}}_0 \cdot  \hat{\boldsymbol{B}}$ is the cosine of the angle between the injected propagation direction $\hat{\boldsymbol{k}}_0$ and the magnetic field direction $\hat{\boldsymbol{B}}$.  Accordingly, this injection distribution rotates as \teq{\theta_{\rm B}} changes across the stellar surface, and is only tied to the zenith direction at the magnetic pole.  The factor  $\Lambda_{\omega}$ in the denominator on the right-hand side of Equation~(\ref{eq:I_mui}) is included to ensure that the intensity distribution $I(\mu_0)$ is normalized to unity.
The photon redistribution anisotropy $A_{\omega}$ depends quadratically on $\mu_0$:
\begin{equation}
	A_{\omega}(\mu_0) \; = \;  \frac{3}{2} \frac{1+ \mathcal{A}(\omega) \mu_0^2}{3+ \mathcal{A}(\omega)} \quad .
	\label{eq:A_omega}
\end{equation}
This characteristic is realized in the asymptotic limit of high opacity, and was demonstrated both numerically \citep{Barchas-2021-mnras} and analytically \citep{Baring-ApJ-2025}.

In a spherical coordinate system ($r, \theta, \phi$) where the injected photon propagation vector $\hat{\boldsymbol{k}}_0$ is chosen to be along the radial direction, the injected electric field (complex) vector $\boldsymbol{\mathcal{E}}_0$ resides in the $\hat{\theta} - \hat{\phi}$ plane and 
therefore can be decomposed into two angular components via
\begin{equation}
	\boldsymbol{\mathcal{E}}_0 \; =\; \mathcal{E}_{\theta} \hat{\theta} + \mathcal{E}_{\phi} \hat{\phi}
    \qquad \hbox{with}\qquad
	\hat{\phi} \; =\; \frac{ \hat{\boldsymbol{B}} \times \hat{\boldsymbol{k}}_0}{|\hat{\boldsymbol{B}} \times \hat{\boldsymbol{k}}_0|}  \quad , \quad  \hat{\theta}  = \hat{\phi} \times \hat{\boldsymbol{k}}_0 \quad  . 
 \label{eq:E_sph-system}
\end{equation}
In this structure, the azimuthal (polar) unit vector is defined such that it is perpendicular (parallel) to the $\hat{\boldsymbol{k}}_0-\hat{\boldsymbol{B}}$ plane.  Moreover, the parallel ($\parallel$) and perpendicular ($\perp$)  mode amplitudes, $\mathcal{E}_{\theta}$ and $\mathcal{E}_{\phi}$, in Equation~(\ref{eq:E_sph-system}) are respectively associated with the ordinary (O) and extraordinary (X) polarization modes -- terms that are commonly used in the literature.
These two amplitudes of the injected electric field vector can be expressed \citep{Barchas-2021-mnras} in terms of Stokes polarization variables:
\begin{equation}
	\mathcal{E}_{\theta} \; =\; \sqrt{\frac{\Pi_{\omega} + \hat{Q}_{\omega}}{2\Pi_{\omega} }}
    \quad , \quad \mathcal{E}_{\phi} \; =\; \frac{ i\hat{V}_{\omega}}{2\Pi_{\omega}	\mathcal{E}_{\theta}} \quad .
 \label{eq:E_theta_phi}
\end{equation}
In these, $\hat{Q}_{\omega}$ and $\hat{V}_{\omega}$ correspond to the linear and circular polarization, respectively, and $\Pi_{\omega}$ denotes the total polarization degree. The empirical approximations of these quantities obtained for the asymptotic configuration of radiation polarization in the limit of high-opacity Thomson scattering are expressed as:
\begin{equation}
    \hat{Q}_{\omega}(\mu_0) \; =\; \frac{\mathcal{A}(\omega)[\mu_0^2 - 1]}{1+ \mathcal{A}(\omega)\mu_0^2} 
    \quad , \quad 
    \hat{V}_{\omega}(\mu_0) \; =\; \frac{2\mathcal{C}(\omega)\mu_0}{1+ \mathcal{A}(\omega)\mu_0^2} 
    \quad , \quad 
    \Pi_{\omega} \; =\; \sqrt{ \hat{Q}_{\omega}^2 + \hat{V}_{\omega}^2 } \quad .
\label{eq:Stokes-inj}
\end{equation}
The coefficients $\mathcal{A}(\omega)$ and $\mathcal{C} (\omega)$ in Equations~(\ref{eq:A_omega}) and (\ref{eq:Stokes-inj}) are related to the anisotropy and circular polarization and depend only  on the photon frequency ratio $\omega /\wcyc$.

In this work, we employ the empirical expressions for $\mathcal{A}$ and $\mathcal{C}$ as given by Equations~(35) and (36) of \citet{Baring-ApJ-2025}. These equations represent an improvement over the versions presented in Equations~(26) and (29) of \citet{Barchas-2021-mnras}, which were also employed in \citet{Hu-2022-ApJ}.
Specifically, in \citet{Barchas-2021-mnras}, the expressions of $\mathcal{A}$ and $\mathcal{C}$ were derived from a numerical fit that was  limited to the frequency range $0.1 \leq \omega / \wcyc \leq 10$. Consequently, they do not accurately capture the correct behaviors outside of this range, particularly in the domain where  $\omega / \wcyc \ll 1$ or $\omega / \wcyc \gg 1$. 
This limitation has been addressed in the recent work by \citet{Baring-ApJ-2025}, where the authors developed the empirical expressions for $\mathcal{A}$ and $\mathcal{C}$ that accurately reproduce the analytic forms of anisotropy and circularity in the extreme frequency regimes. This improvement makes the simulation more expedient for studying neutron stars of various magnetizations, from low-field ($\omega / \wcyc \gg 1)$ to high-field ($\omega / \wcyc \ll 1$) ones. The {\sl MAGTHOMSCATT} simulation was validated in \citet{Barchas-2021-mnras} for both intensity and polarization properties by direct comparisons with results obtained at various frequency ratios, $\omega/\wcyc = 0.25, 0.5, 2, 10$,  by \cite{Whitney-1991-ApJS} and the numerical solutions to non-magnetic radiative transfer integro-differential equation by \citet{Sunyaev-1985-AandA}. In the latter case, where $\omega /\wcyc \gg 1$, the magnetic Thomson cross section reduces to the familiar Thomson cross section $\sigt$, and the magnetic field has negligible impact on the scattering process.

The detailed numerical steps for implementing the AP injection protocols in the {\sl MAGTHOMSCATT} simulation are outlined in Section 2.2 of \citet{Hu-2022-ApJ}. When the convergent configuration is achieved, the resulting intensity and polarization are insensitive to the injection protocols.

An important parameter characterizing the radiation emergent from the atmospheric slabs is the effective optical depth parameter $\tau_{\rm eff}$, which depends on the magnetic field direction and relates to the Thomson optical depth \teq{\taut} via (see Equation~(5) in \citet{Hu-2022-ApJ}):
\begin{equation}
   \tau_{\rm eff} \; =\; \frac{\taut}{2} \left[\sin^2 \theta_{\rm B} 
   + \Sigma_{\rm B} \left(1+ \cos^2 \theta_{\rm B}\right)\right] 
   \quad ,\quad
   \taut \; =\; n_e \sigt h  \quad ,
 \label{eq:tau_eff}
\end{equation}
where $h$ is the thickness of the atmosphere layer, and $n_e$ is the electron number density, which is assumed to be uniform within the slab, a choice of convenience not necessity.  The optimal choice for the effective optical depth parameter corresponds to the minimum $\tau_{\rm eff}$ required for the code to reach a convergence configuration.  This is where the scattering is saturated, and the anisotropy and polarization characteristics remain invariant with further increases of $\tau_{\rm eff}$.  The simulation using the AP injection protocol reaches convergence at a somewhat or much lower $\tau_{\rm eff}$ than the IU counterpart, thereby significantly reducing the run times \citep{Hu-2022-ApJ, Baring-ApJ-2025}.  
In particular, Table~3 in \cite{Baring-ApJ-2025} illustrates that the AP injection protocol leads to a substantially faster convergence than does the IU counterpart for implementations in the sub-cyclotronic domain and for non-polar colatitudes. In these cases, the run time for AP is up to three orders of magnitude smaller than that for IU injection.

In addition to the AP injection approach outlined above, we have recently incorporated into the simulation a new protocol aiming to making the code more efficient in the  $\omega / \wcyc \ll 1$ regime. This upgrade will be introduced in Section~\ref{sec:new_protocol}. Since the vintage of the code presented in \citet{Hu-2022-ApJ},  {\sl MAGTHOMSCATT} has been parallelized using OpenMP and Message Passing Interface (MPI) and is being run on Rice University’s NOTS cluster\footnote{https://researchcomputing.rice.edu/rice-supercomputing-nots}.

\subsection{Photon diffusion in atmospheric slabs}
 \label{subsec:diffusion}

In this Section, we examine spatial distributions of photons, aiming to understanding the main characteristics of the diffusion of photons, within an atmospheric slab.  To sample the distance $s$ traveled by a photon between two consecutive scattering events within an atmospheric slab, {\sl MAGTHOMSCATT} uses standard Poisson statistics as in \citet{Barchas-2021-mnras}:
\begin{equation}
	s \; = \; s (\omega, \mu_i) \; = \; -\frac{\log\xi_s}{n_e \sigma(\omega,\mu_i)} \quad ,
\end{equation}
in which $\xi_s$ is a uniform random variate on the interval $[0,1]$. As already mentioned, the total magnetic Thomson scattering cross section $\sigma(\omega, \mu_i)$ is dependent on the frequency $\omega$  of the photon and its incident propagation direction $\hat{\boldsymbol{k}}_i$ with respect to the magnetic field direction $\hat{\boldsymbol{B}}$; see Equation~(9) in \citet{Barchas-2021-mnras}. The distance $s_0$ that a photon travels before its first scattering after being injected at the base of the slab is $s_0 = s(\omega , \mu_0)$, with $\mu_0$ being sampled using the AP injection protocol according to Equation~(\ref{eq:I_mui}).

\begin{figure}[t]
	\centering
	\includegraphics[width=0.45\linewidth]
   % {3dtrajectories.pdf}
   {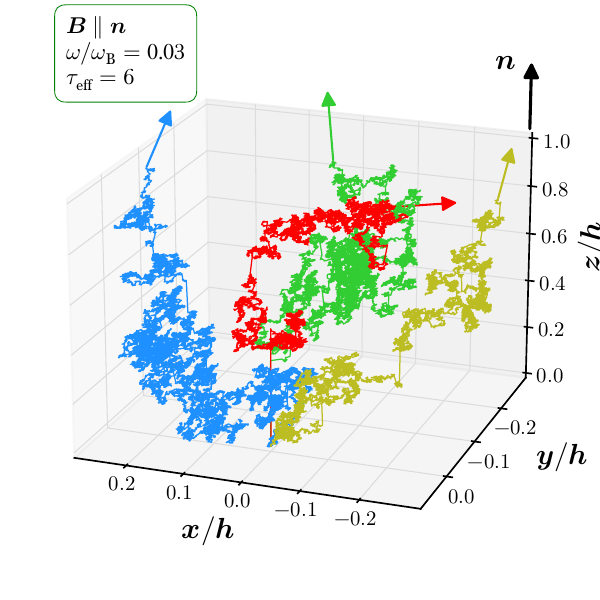}
   \hspace{5pt}
    \includegraphics[width=0.52\linewidth]
    %{2d-projection.pdf}
    {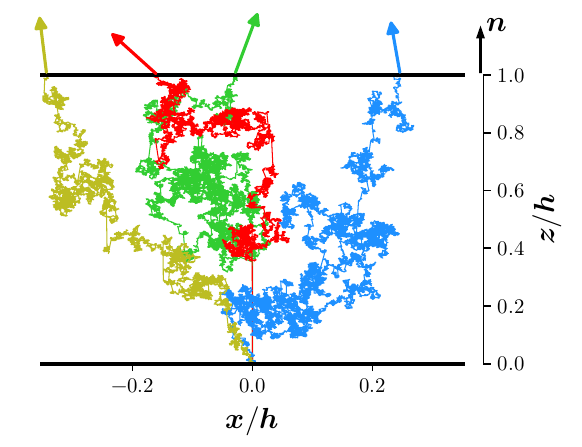}
	\caption{Left: 3D trajectories of four photons in the {\sl MAGTHOMSCATT} simulation of an atmospheric slab of height $h = \tau_T / (n_e\sigt)$ at the magnetic pole ($\boldsymbol{B} \parallel \boldsymbol{n}$) . The $z$-axis (zenith) is oriented along the direction of the normal vector $\boldsymbol{n}$. The green, red, blue, and olive trajectories correspond to photons that scatter 1068475, 614756,  1299544, and 681861 times, respectively, before exiting through the upper boundary of the slab. Right: projection of these trajectories onto the $x-z$ plane. The arrows indicate the photon propagation directions after emerging from the upper boundary into the magnetosphere. The simulation was performed at a frequency ratio of $\omega / \wcyc = 0.03$ and an effective optical depth of $\tau_{\rm eff} = 6$, as defined in Equation~(\ref{eq:tau_eff}).  At injection, the photons with red and green trajectories have $\perp$-polarization mode, while those with blue and olive trajectories were injected in $\parallel$ mode.  Upon emergence, they are all in $\parallel$-polarization mode.}
	\label{fig:trajectories}
\end{figure}

After each scattering event, the photon emerges in a  direction $\hat{\boldsymbol{k}}_f$ that is determined by the polarization-dependent differential cross section using the accept-reject method; see Equation~(19) in \citet{Barchas-2021-mnras}. The photon electric field vector then can be calculated using Equation~(6) therein. Accordingly, this formalism allows for tracking the complete polarization characteristics of each photon.  The photon position where the $j$th scattering event occurs reads
\begin{equation}
 	\boldsymbol{r}_{j} \; = \; \boldsymbol{r}_{j-1} + s(\omega, \mu_{i,j}) \hat{\boldsymbol{k}}_{i, j} \; = \; \boldsymbol{r}_{j-1} + s(\omega, \mu_{f,j-1})\hat{\boldsymbol{k}}_{f, j-1} \quad ,
 \label{eq:displacement}
\end{equation}	
wherein it is implicit that $\boldsymbol{k}_{f, j-1} = \boldsymbol{k}_{i,j}$.
In a Cartesian coordinate system with the $z$-axis being along the normal vector $\boldsymbol{n}$ of the slab and the origin being located at the injection point, the condition for exiting the slab of thickness $h$ at the top is $z = \boldsymbol{r} \cdot \hat{\boldsymbol{n}} > h$, while that for exiting at the bottom is $z = \boldsymbol{r} \cdot \hat{\boldsymbol{n}} < 0$.

\begin{figure}[t]
    \centering
    \includegraphics[width=1\linewidth]{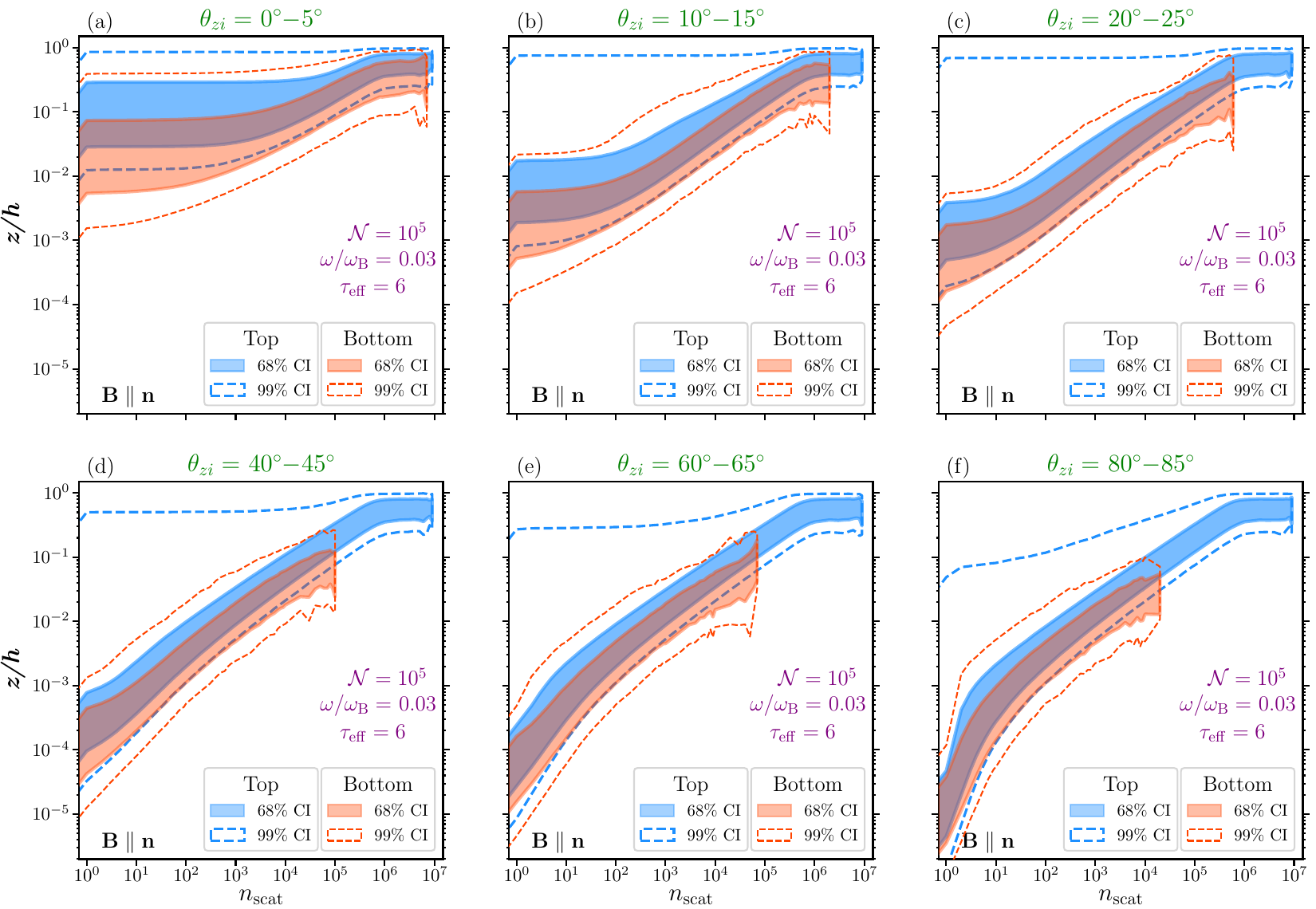}
    \caption{The phase space diagram of vertical position versus scattering number.  $68\%$ (solid regions) and $99\%$ (dashed regions) confidence intervals (CIs) of vertical positions $z/h$ for $\mathcal{N} = 10^{5}$ photons recorded at the top (blue) and bottom (red-brown) of the slab as a function of the scattering number $n_{\rm scat}$. Six restrictive ranges of zenith angles were considered: $\theta_{zi} \in [0^{\circ}, 5^{\circ}]$ (panel a),  $\theta_{zi} \in[10^{\circ}, 15^{\circ}]$ (panel b),  $\theta_{zi} \in[20^{\circ}, 25^{\circ}]$ (panel c),  $\theta_{zi} \in[40^{\circ}, 45^{\circ}]$ (panel d), $\theta_{zi} \in[60^{\circ}, 65^{\circ}]$ (panel e), and  $\theta_{zi} \in[80^{\circ}, 85^{\circ}]$ (panel f).  The simulations were performed at $\theta_{\rm B} = 0^{\circ}$ (i.e., at the magnetic pole), $\omega / \wcyc = 0.03$, and $\tau_{\rm eff} = 6$.}
    \label{fig:68-99CI-band}
\end{figure}

For illustrative purposes, we performed  simulations for an atmospheric slab located at the magnetic pole ($\theta_{\rm B} = 0^{\circ}$) for a photon frequency of  $\omega /\wcyc = 0.03$. The effective optical depth is fixed at $\tau_{\rm eff} = 6$. We only accounted for photons that scatter at least once.
In Figure~\ref{fig:trajectories}, we present the 3D trajectories (left panel) of four selected photons, along with their projections onto the $x-z$ plane (right panel).   Among these, the photons represented by the red and green trajectories were injected along the slab normal, while the blue and olive ones were respectively injected at $\theta_{zi} = 40^{\circ}$ and $\theta_{zi} = 80^{\circ}$, where $\theta_{zi} = \arccos{(\hat{\boldsymbol{k}}_0\cdot \hat{\boldsymbol{n}})}$. It is evident from this Figure that the photons injected along the zenith travel  a significant distance before their first collision. In particular, the first scattering event of the red photon occurs at $z = 0.48h$, and at $0.35h$ for the green photon. In contrast, for the photons with blue and olive traces, their first scattering events happen very close to the base, at $z =  8 \times 10^{-5} h$ and $z = 3 \times 10^{-6} h $, respectively.  Moreover, we can observe from Figure~\ref{fig:trajectories} that whenever a photon propagates close to the field direction, it covers a longer distance before its subsequent scattering than when it travels  obliquely to the field. This property results from the reduction in the cross section within the MSC in the $\omega / \wcyc \ll 1$ domain, as discussed in Section~\ref{subsec:magThomson}.
Regardless of the injection angles, all these individual photons undergo a substantial number of scatterings before exiting, indicating that the convergence configuration is expected to be established with the chosen effective optical depth.
Note that while the example photons were injected by sampling the anisotropy/polarization protocol identified in Equation~(\ref{eq:Stokes-inj}), in practice they neatly divided into two approximately linear polarizations in the $\omega / \wcyc \ll 1$ domain, as noted in the figure caption.

To quantitatively assess the evolution of photon vertical positions within the slab { before they escape}, in Figure~\ref{fig:68-99CI-band}, we display the $68\%$ (solid regions) and $99\%$ (dashed regions) confidence intervals (CIs) of $z/h$  distributions for $\mathcal{N} = 10^{5}$ photons collected at the top (shown in blue) and bottom (shown in red-brown) of the slab as a function of the scattering number $n_{\rm scat}$. For this investigation, we injected photons within six restrictive ranges of zenith angles: $\theta_{zi} \in [0^{\circ}, 5^{\circ}]$~(panel a),  $\theta_{zi} \in[10^{\circ}, 15^{\circ}]$~(panel b),  $\theta_{zi} \in[20^{\circ}, 25^{\circ}]$~(panel c),  $\theta_{zi} \in[40^{\circ}, 45^{\circ}]$~(panel d), $\theta_{zi} \in[60^{\circ}, 65^{\circ}]$~(panel e), and  $\theta_{zi} \in[80^{\circ}, 85^{\circ}]$~(panel f).    In general, it is evident that the distributions of top and bottom-escaping photons largely overlap, a consequence of the Markovian nature of the diffusion.  For this reason, it proves difficult to isolate top-escaping photons at early stages of their diffusion history. The probabilities of photons escaping from the top in panels (a) through (f) are: $5.44\%$, $0.60\%$, $0.24\%$, $0.10\%$, $0.06\%$, and $0.04\%$. Since the vast majority of injected photons escape through the bottom of the slab, such a top/bottom separation facility would have the potential to improve the simulation speed.  
In each of these six panels, the distributions were obtained for $10^5$ photons escaping from the top and $10^5$ photons escaping from the bottom, injected sufficiently large numbers of photons so as to ensure identical statistics for the two exit populations.
{We note that once a photon exits the slab, either through the top or the bottom, its position is no longer included in the $z/h$ distribution of photons within the slab. As the number of scatterings, $n_{\rm scat}$, increases, the number of photons remaining within the slab decreases. The right edges of the CI regions in Figure~\ref{fig:68-99CI-band} correspond to the $n_{\rm scat}$ values  at which fewer than 10 photons remain in the slab, that is, $0.01\%$ of the total collected photons.}

The initial scattering event of photons, whether they exit from the top or the bottom of the slab, transpires at higher $z$ for smaller values of $\theta_{zi}$.  This feature is consistent with the trajectory behavior exhibited in Figure~\ref{fig:trajectories}.  As one might expect, photons injected at smaller angles $\theta_{zi}$ penetrate deeper into the slab on average, and hence experience more scattering events overall {if they eventually escape through the lower boundary}.  For all injection zenith angles, the total number of scatterings for photons exiting the top of the slab always exceeds that for those exiting through the bottom.  Yet, when $\theta_{zi} < 25^\circ$ (see panels a, b, c), photons scatter many times, in excess of $n_{\rm scat, max} > 10^6$,  before they can exit from the bottom (shown in red-brown). In contrast, photons injected at larger $\theta_{zi}$ angles are more likely to escape through the bottom after significantly fewer scatterings, as these events tend to happen closer to the base of the slab.  {This also explains why the CI regions for bottom-escaping photons terminate at lower $n_{\rm scat}$ for larger $\theta_{zi}$ angles.} Regardless of the injection angles, as the number of scatterings increases, photons diffuse, and their distribution spreads to higher $z/h$. When $n_{\rm scat}$ is small, the majority of photons are still concentrated near the base, and therefore, the probability of them escaping the slab from the lower boundary is high.  As photons  continue to undergo more scattering events, a portion of photons exit from the bottom, while the rest continue to slowly move upward, resulting in higher $z/h$.  At large $n_{\rm scat}$,  the remaining photons are those that have penetrated sufficiently deeply into the slab. Consequently,  they  have lower probabilities of exiting the slab through the bottom. 

For photons emerging from the top (shown in blue), the average numbers of scatterings in panels (a) through (f) are as follows:  $1.56 \times 10^6$, $1.62 \times 10^6$, $1.63 \times 10^6$, $1.64 \times 10^6$, $1.64 \times 10^6$, and $1.64 \times 10^6$, i.e., exhibiting only a small variation with $\theta_{zi}$. This result implies that a small fraction ($\sim 5$\%) of photons injected at  $\theta_{zi} < 5^{\circ}$ escape the upper boundary with fewer scatterings, and that the scattering number does not change significantly when $\theta_{zi} > 20 ^{\circ}$. The maximum number of scatterings, $n_{\rm scat, max} \sim 10^7$, shows no visible dependence on $\theta_{zi}$.  Moreover, the blue bands in all panels appear very similar when $n_{\rm scat} > 10^5$. This indicates that the number of scatterings in our Monte Carlo simulation is sufficiently large that a truly Markovian domain has been achieved, and all memory of the injection anisotropy and polarization distributions is lost.

\begin{figure}[t]
	\centering
	\includegraphics[width=1\linewidth]{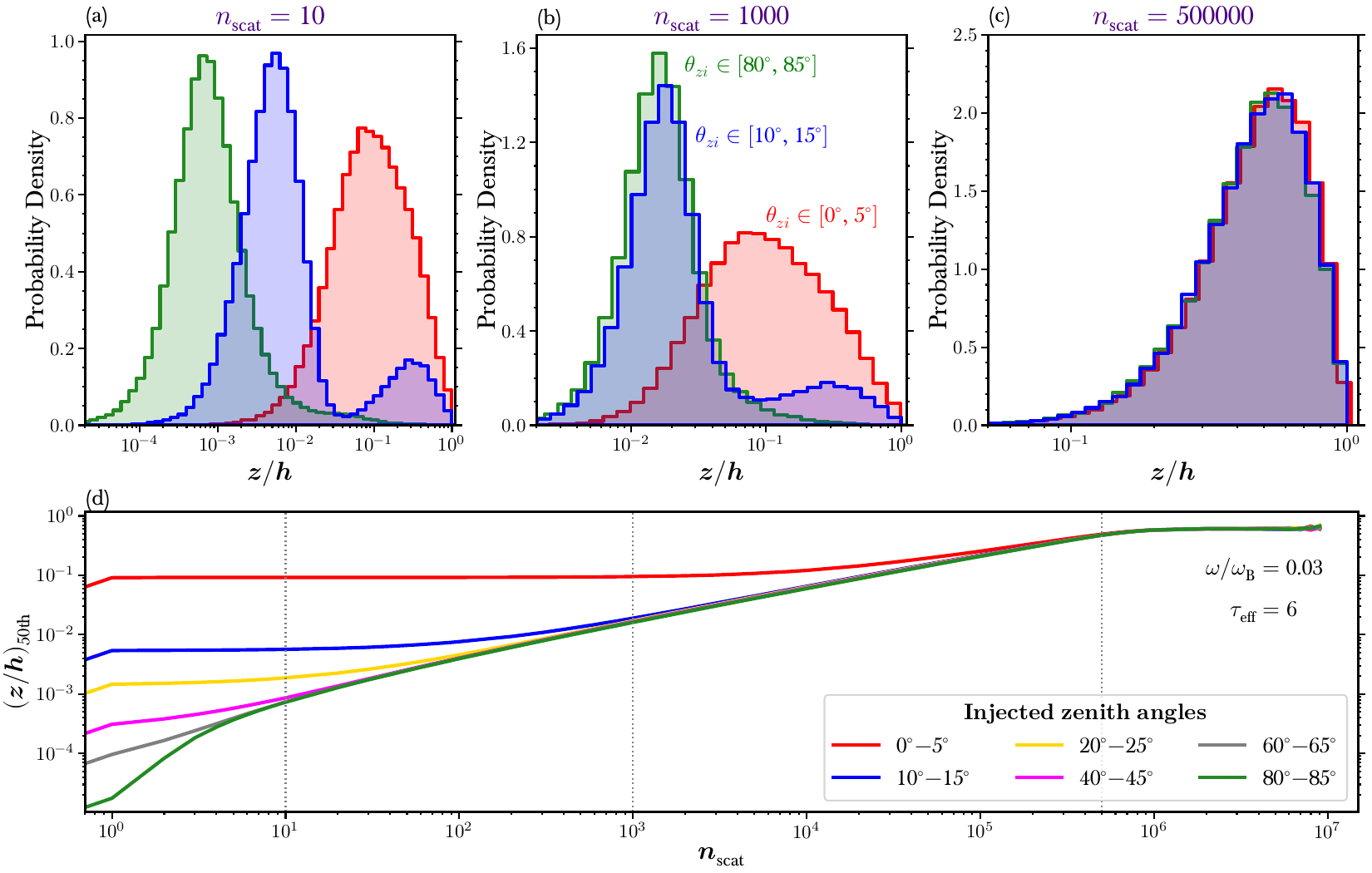}
	\caption{Top panels: Distributions of $z/h$ of photons exiting the slab from the top at $n_{\rm scat}$ = 10 (panel a), $10^3$ (panel b), and $5 \times 10^5$ (panel c) for $\theta_{zi} \in [0^{\circ}, 5^{\circ}]$ (red), $\theta_{zi} \in [10^{\circ}, 15^{\circ}]$ (blue), and $\theta_{zi} \in [80^{\circ}, 85^{\circ}]$ (green).  Bottom panel: median values of $z/h$ for top photons injected in six $\theta_{zi}$ ranges. The simulations were performed at $\theta_{\rm B} = 0^{\circ}$, $\omega / \wcyc = 0.03$, and $\tau_{\rm eff} = 6$. }
	\label{fig:distribution}
\end{figure}

To elaborate further, in Figure~\ref{fig:distribution}, we plot the distributions of photons exiting the slab \underline{from the top} at different scattering numbers, $n_{\rm scat} = 10$ (panel a), $n_{\rm scat} = 10^3$ (panel b), and $n_{\rm scat} = 5 \times 10^5$ (panel c), for three selected ranges of zenith angles at injection, namely, $\theta_{zi} \in [0^{\circ}, 5^{\circ}]$ (red), $\theta_{zi} \in [10^{\circ}, 15^{\circ}]$ (blue), and $\theta_{zi} \in [80^{\circ}, 85^{\circ}]$ (green). These distributions constitute vertical slices of the blue swaths in three of the panels in Figure~\ref{fig:68-99CI-band}; they are normalized such that the areas under the distributions equal unity.  In panel (d), the medians of the $z/h$ distributions of top-escaping photons across all six considered $\theta_{zi}$ ranges are compared.
This lower panel illustrates that the median values differ from one another at low \(n_{\rm scat}\), with the median decreasing as \(\theta_{zi}\) increases. These curves begin to merge as the number of scatterings increases.  When $n_{\rm scat}$ reaches $ \sim 5\times 10^5$, they merge completely. 
This behavior is reflected in the probability density distributions in the top row, where the three distributions entirely coincide in panel (c). 
Thus, after $\sim 5\times 10^5$ scatterings, the spatial distributions of photons within the slabs become independent of $\theta_{zi}$, meaning they lose any residual memory of their injection.  This demonstrates that the diffusion in the slab has reached the Markovian domain, for which the anisotropy and polarization distributions are fully described by the analysis in \cite{Baring-ApJ-2025}, and for which the emergent intensity and polarization characteristics are insensitive to specific initial conditions at injection and will not be impacted by further scatterings.

\section{Emergent Anisotropy and Polarization Properties}
\label{sec:emergent}

The focus of this Section is a survey of the emergent anisotropy and polarization properties from local atmospheres in $\omega / \wcyc \ll 1$, $\omega / \wcyc \sim 1$ (Sections~\ref{subsec:magnetar} and \ref{subsec:polarization_diversity}) and $\omega / \wcyc \gg 1$ domains (Section~\ref{subsec:weak-field}), exploring these in greater depth than was offered in \cite{Barchas-2021-mnras} and \cite{Hu-2022-ApJ}.  One agenda motivating such a survey is in identifying paths to accelerate the run efficiency of the {\sl MAGTHOMSCATT} simulation.

In presenting simulation results for localized atmospheres, it is convenient to choose the normal vector  of the slab, $\hat{\boldsymbol{n}}$, as the reference direction.  For any symmetric magnetic field configuration, such as a dipole in the Schwarzschild metric \citep{Wasserman1983}, or axisymmetric twisted field morphologies in flat \citep{Thompson-2002-ApJ} and Schwarzschild \citep{Hu-2022-ApJ-opac} spacetimes, at the magnetic pole $\hat{\boldsymbol{n}} \parallel\hat{\boldsymbol{B}} $, while $\hat{\boldsymbol{n}} \perp \hat{\boldsymbol{B}} $ at the equator.
For each photon that manages to exit the slab from the top, we recorded its injected zenith angle $\theta_{zi} = \arccos{(\hat{\boldsymbol{k}}_0\cdot \hat{\boldsymbol{n}})}$, with $\hat{\boldsymbol{k}}_0$ being the injected propagation vector, the number of scatterings the photon undergoes $n_{\rm scat}$, the emergent complex electric field vector $\boldsymbol{\mathcal{E}}_{\rm S}$,  and its emerging zenith angle $\theta_z = \arccos{(\hat{\boldsymbol{k}}_{\rm S}\cdot \hat{\boldsymbol{n}})}$, where $\hat{\boldsymbol{k}}_{\rm S}$ is the unit wavevector of the photon emerging at the neutron star surface. With the electric field vector, the Stokes parameters ($I$, $Q$, $U$, $V$) -- where $I$ denotes the radiation intensity, $Q$ and $U$ carry information on the linear polarization degree and angle, and $V$ represents the circular polarization -- can be calculated using Equation~(A4) in \citet{Hu-2022-ApJ} with the following coordinate system:
\begin{equation}
	\hat{z}_{\rm S} \; =\; \hat{\boldsymbol{k}}_{\rm S} \ \ , \ \ \hat{y}_{\rm S} \; =\;  \frac{ \hat{\boldsymbol{n}} \times \hat{\boldsymbol{k}}_{\rm S} }{|\hat{\boldsymbol{n}} \times \hat{\boldsymbol{k}}_{\rm S}|}   
    \quad , \quad \hat{x}_{\rm S} \; =\; \hat{y} \times \hat{\boldsymbol{k}}_{\rm S} \quad . 
 \label{eq:xyz_S}
\end{equation}
Due to the azimuthal symmetry, an integration over azimuthal angles around the zenith yields $U = 0$ 
locally; this is a choice that is made for Figure~\ref{fig:slabs-1d} here, and for a number of illustrations in \citet{Barchas-2021-mnras}.

\subsection{Distributions for the $\omega / \wcyc \ll 1$ and $\omega / \wcyc \sim 1$ domains}
 \label{subsec:magnetar}
\begin{figure}[t]
	\centering
	\includegraphics[width=1.0\linewidth]{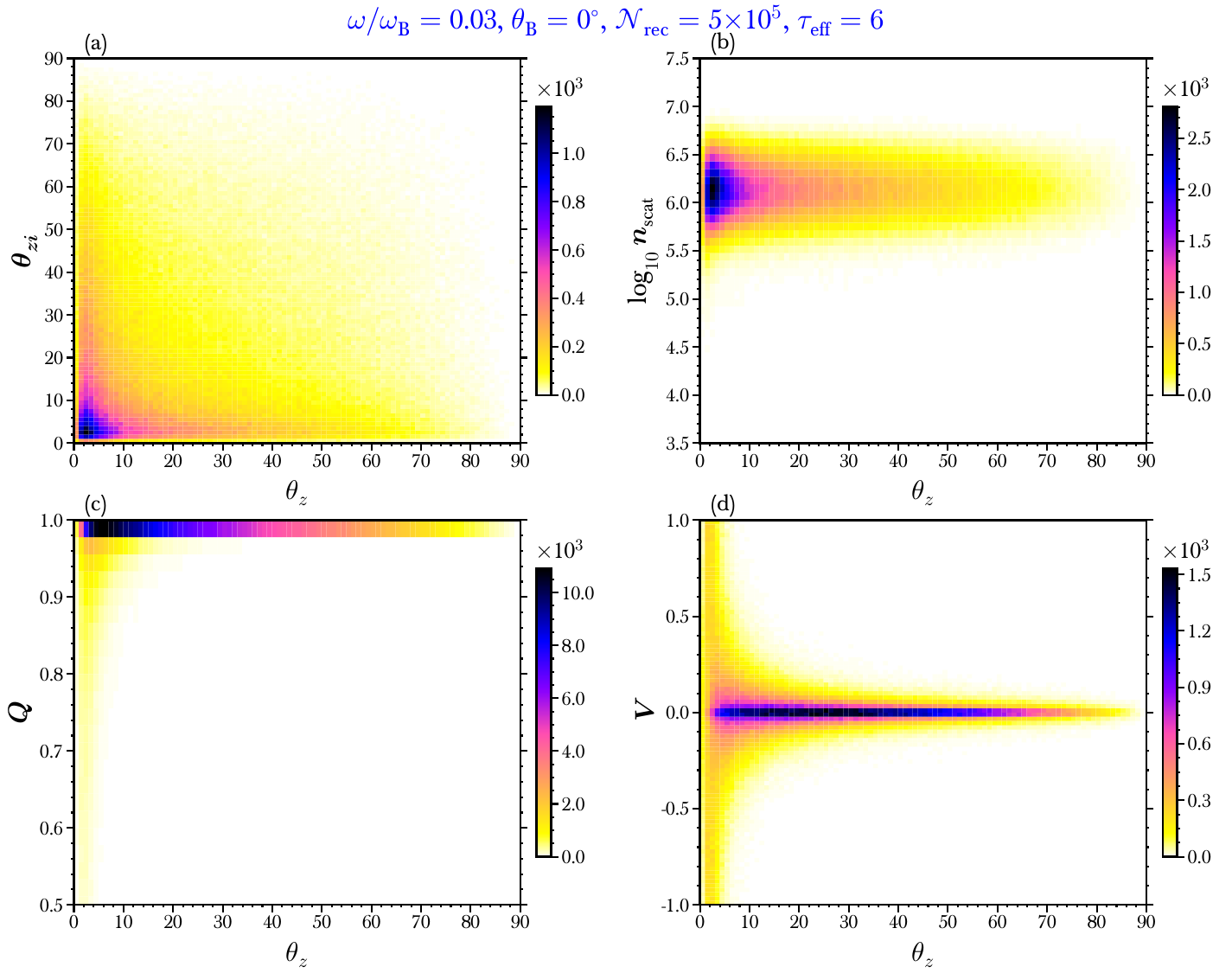}
	\caption{Distributions of photons exiting the top of an atmospheric slab positioned at the magnetic pole ($\theta_{\rm B} = 0^{\circ}$).  These density maps are functions of the emerging zenith angle $\theta_z$ (x-axes) and the injected zenith angle $\theta_{zi}$ (panel a), logarithmic scattering numbers $\log_{10}{n_{\rm scat}}$ (panel b), Stokes parameter $Q$ (panel c), and Stokes parameter $V$ (panel d).  The maps are for $\omega / \wcyc = 0.03$, appropriate for surface X-ray emission from a young pulsar.  The colorbar on the right of each panel indicates the number of photons in each bin, so that the sum of these over all pixels equals the total number of photons recorded is $\mathcal{N}_{\rm rec} = 5\times 10^5$.  The pixel size is 1 degree for both \teq{\theta_z} and \teq{\theta_{zi}}}.
	\label{fig:fig-slab-2d-theta}
\end{figure}

In Figure~\ref{fig:fig-slab-2d-theta}, we display 
various two-dimensional distributions pertaining to the photon population as functions of of the emergent zenith angle $\theta_z$, specifically for a field directed parallel to the local zenith direction ($\theta_{\rm B} = 0^{\circ}$), i.e., at the magnetic pole.  These are the injected zenith angle $\theta_{zi}$  (panel a), the logarithm of the number of scatterings  $\log_{10} n_{\rm scat}$ a photon experiences before exiting the upper boundary (panel b), Stokes $Q$ (panel c), and Stokes $V$ (panel d).  These results were for $\mathcal{N}_{\rm rec} = 5\times 10^5$ photons escaping from upper boundary of an atmospheric slab, all of which were scattered at least once, and were injected using the {\bf B} field-relative anisotropy and polarization distributions defined by Equations~(\ref{eq:A_omega}), (\ref{eq:E_theta_phi}) and~(\ref{eq:Stokes-inj}).  The photon frequency ratio is fixed at $\omega /\wcyc = 0.03$,  a value that corresponds to 1 keV emission for a surface polar field of \teq{B\sim 2.7 \times 10^{12}}Gauss, around 40\% of the polar for the Vela pulsar.  The convergence configuration is achieved with $\tau_{\rm eff} =6$, or, equivalently, $\taut = \tau_{\rm eff} / \Sigma_{\rm B} \approx 6.6 \times 10^3$. 

As shown in panel (a), the majority of final photons are those propagating along the magnetic field lines, $\theta_z \lesssim 4^{\circ}$, primarily originating from small injected zenith angles $\theta_{zi}$.  This suppression of photons traveling obliquely to the field occurs due to the dominance of the $\parallel \rightarrow \parallel$ cross section at large incident angles, as evident in Equation~(\ref{eq:dsig_magThom}). The large cross section and the preference for significant to large scattering angles make it unlikely for these photons to ever reach the upper boundary, causing most of them to scatter multiple times and eventually exit from the bottom of the slab. Consequently, the overall probability of eventual escape through the upper surface is very low, only around $0.2 \%$.  This highlights the inherent inefficiency of the simulation when \teq{\theta_{\rm B}=0^{\circ}} \underline{and} $\omega /\wcyc \ll 1$, a limitation that will be mitigated somewhat via simulation refinements, as discussed in Section~\ref{sec:new_protocol} below. 

In panel (b) of Figure~\ref{fig:fig-slab-2d-theta}, the distribution of the number of scatterings with emergent zenith angle is depicted.  This provides a sense of the coupling between diffusion interior to the slab and escape at the upper surface.  On average, each emerging photon undergoes approximately $1.6 \times 10^6$ scatterings. Close to the zenith direction, i.e., when $\theta_z \lesssim 4^{\circ}$, there is a small population of photons that go through substantially fewer scatterings, $n_{\rm scat} \approx 10^5$. This is due to the reduction of the cross section along this direction, where the cross sections of photons in the two polarization modes, $\sigma_{\perp}$ and $\sigma_{\parallel}$, are of the same order of magnitude and proportional to $\sigt \Sigma_{\rm B} \propto \sigt (\omega / \wcyc )^2$.  At the optical depth we employed, $\tau_{\rm eff} = 6$, there is a small number of photons exiting the surface without scattering subsequent to injection, constituting only about $0.2\%$ of those traveling through the slab upper boundary; these unscattered photons are not included in the final emergent distributions.  For atmospheres of higher optical depths, photons are unlikely to propagate freely through the surface layer without any collision. Indeed, we have checked that at a sufficiently sizable optical depth $\tau_{\rm eff} =  50$ (for $\omega / \wcyc = 0.03$ at $\theta_{\rm B} = 0 ^{\circ}$), the contributions from unscattered photons are completely eliminated, while the final angular distributions of intensity and polarization remain unchanged. This is because the scattering numbers at $\tau_{\rm eff} \gtrsim 6$ are sufficiently large that the simulation has achieved a Markovian domain, where the amount of diffusion in scatterings has completely obscured the information about the injection, and further scattering does not impact the emergent radiation configuration.  Accordingly, it is much more efficient to perform the simulation at a moderate optical depth and consider only scattered photons when exhibiting simulation output.

The distribution maps of Stokes parameters $Q$ and $V$ with emergent zenith angle are respectively shown in panels (c) and (d) in Figure~\ref{fig:fig-slab-2d-theta}.  Due to the choice of coordinates, $U \approx 0$ at $\theta_{\rm B} = 0^{\circ}$ for sub-cyclotronic frequencies, and is therefore not displayed. The dominance of the $\parallel \rightarrow \parallel$ cross section results in the generation of $\parallel$-mode photons being favored.  Consequently, the emerging radiation is highly linearly polarized in directions away from the magnetic field, with $Q\sim 1$ and $V \sim 0$ for $\theta_z \gtrsim 2 \theta_{\rm MSC} \sim 4^{\circ}$, i.e., outside the magnetic scattering cone.  Inside this cone, linear depolarization naturally occurs as the relative disparity in differential cross sections for the different polarization transition modes becomes more muted; then the emergent signal possesses a broader distribution with \teq{Q\sim 0.5 - 1} and \teq{V} sampling a broad range of values \teq{-1 \leq V \leq 1}. The summation of $V$ over all photons in each $\theta_z$ bin in panels (d) gives $V/I \approx 0.025$ along the magnetic field direction (i.e., for \teq{0 \leq \theta_z \leq 1^{\circ}}), indicating that photons with positive $V$ are slightly more favored.  Note that while the range of \teq{Q} is truncated to \teq{[0.5,1]} in panel (c) to enhance visual clarity, there are still a small number of photons that possess \teq{-1 < Q < 0.5} when \teq{\theta_z < \theta_{\rm MSC}}.  

\begin{figure}[t]
	\centering
	\includegraphics[width=1.0\linewidth]{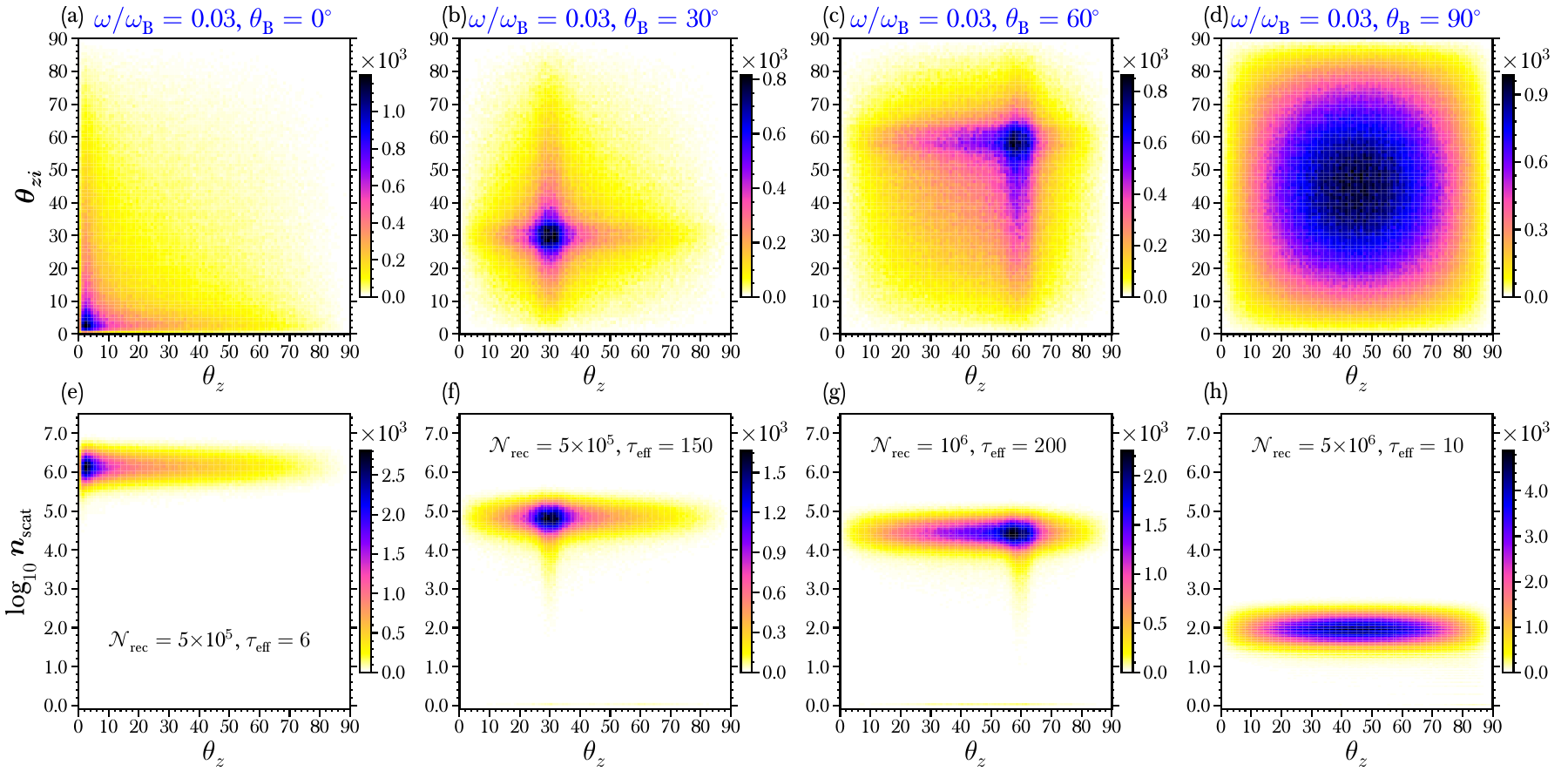}
	\caption{Distributions of photons exiting the top of the slab as a function of the emerging zenith angle $\theta_z$ and: the injected zenith angle $\theta_{zi}$ (top row) and logarithmic scattering numbers $\log_{10}{n_{\rm scat}}$ (bottom row) for $\omega / \wcyc = 0.03$. Four columns correspond to four different field orientations, $\theta_{\rm B} = 0^{\circ}, 30^{\circ},  60^{\circ}, 90^{\circ}$, corresponding to surface locales spanning the pole to the equator.  The total number of photons recorded $\mathcal{N}_{\rm rec}$ and the value of the effective optical depth $\tau_{\rm eff}$ for each $\theta_{\rm B}$ are indicated in the bottom panel of each column. The colorbar on the right of each panel indicates the number of photons in each bin.}
	\label{fig:fig-slab-2d-theta-4thB}
\end{figure}

One advantage of the {\sl MAGTHOMSCATT} simulation is that it can be employed for any magnetic colatitude since it routinely accommodates arbitrary orientations of the local magnetic field. Therefore, we extended the above analysis to nonpolar magnetic colatitudes,  specifically at $\theta_{\rm B} =  30^{\circ}$, $60^{\circ}$, and  $90^{\circ}$, and compared the resulting  $\theta_{zi}-\theta_{z}$ (top row) and $\log_{10}n_{\rm scat}-\theta_{z}$ (bottom row) distributions with those obtained at $\theta_{\rm B} = 0^{\circ}$ in Figure~\ref{fig:fig-slab-2d-theta-4thB}. Similar to the polar case, at $\theta_{\rm B} = 30^{\circ}$ (panel b) and  $\theta_{\rm B} = 60^{\circ}$ (panel c), photons are beamed around the magnetic field direction, although the collimation is less pronounced, in part because at each \teq{\theta_z} there are many azimuths that contribute to the output displayed.  At the equator (panel d), the beaming effect is completely washed out, and the photon distribution is broad and has lost any association with the field direction.  We note that the output displayed in Figure~\ref{fig:fig-slab-2d-theta-4thB} does not reflect intensities exiting the slab, just photon counts.  Accordingly the counts are missing the solid angle and flux weighting correction factors that are adopted in \cite{Hu-2022-ApJ} and in Figure~\ref{fig:slabs-1d} below. In the \teq{\theta_{\rm B} = 90^{\circ}} illustration (panel d), since the distribution is isotropic everywhere except near and in the MSC, the zenith angle dependence is then simply \teq{\propto \sin \theta_z\cos\theta_z}, and similarly for the injection dependence on \teq{\theta_{zi}}, thereby generating the peaks discernible at \teq{45^{\circ}} in the distributions for both angles.

The scattering numbers (lower row in Figure~\ref{fig:fig-slab-2d-theta-4thB}) required for convergence significantly drop from $\sim 10^6$ scatterings at the pole to only $\sim 10^2$ at the equator. This decrease arises because the injected intensity and polarization distributions in the latter case closely resemble the emergent ones, both being close to isotropy, as will be shown in Figure~\ref{fig:slabs-1d}. As a result, the convergence configuration is realized only after about $\sim 100$ scatterings. This highlights the efficacy of the AP injection protocol, with which the run time could be reduced by about three orders of magnitude for the equatorial case in the $\omega / \wcyc \ll 1 $ domain \citep{Baring-ApJ-2025}.

{We found that the average number of scatterings, $\langle n_{\rm scat} \rangle$,  of emergent photons in the simulations in Figure~\ref{fig:fig-slab-2d-theta-4thB}  increases approximately quadratically with the Thomson optical depth, $\langle n_{\rm scat} \rangle \propto \taut^2$. From  Equation~(\ref{eq:tau_eff}), it follows that $\langle n_{\rm scat} \rangle$ also grows with $\tau_{\rm eff}^2$. The $\langle n_{\rm scat} \rangle-\taut^2$ relation is determined by the magnetic scattering cross sections, which vary with the photon’s polarization states and their propagation directions with respect to the magnetic field vector; see Equation~(\ref{eq:dsig_magThom}). In the considered cases, the magnetic Thomson cross sections are reduced, particularly along the field direction,  compared to the non-magnetic one. Therefore, the slopes of the $\langle n_{\rm scat} \rangle-\taut^2$ relations in these simulations are less than one. For the four field orientations, $\theta_{\rm B} = 0^{\circ}, 30^{\circ},  60^{\circ}, 90^{\circ}$, the respective relation between the average scattering number and the Thomson optical depth are: $\langle n_{\rm scat}  \rangle \approx 0.037 \taut^2$, $\langle n_{\rm scat}  \rangle \approx 0.054 \taut^2$, $\langle n_{\rm scat}  \rangle \approx 0.113 \taut^2$, and $\langle n_{\rm scat}  \rangle \approx 0.279 \taut^2$. Since there are more photons traveling obliquely relative to the field direction for larger $\theta_{\rm B}$, the proportionality coefficient increases from the pole to the equator, as expected. We note that this $\taut^2$ dependence holds for sufficiently large $\taut$ values. When $\taut$ is small, such as in the simulation for $\theta_{\rm B} = 90^{\circ}$, an additional correction term linear in $\taut$ is needed to better describe the relationship between $\langle n_{\rm scat} \rangle$ and $\taut$. }

\begin{figure}[ht]
	\centering
	\includegraphics[width=1.0\linewidth]{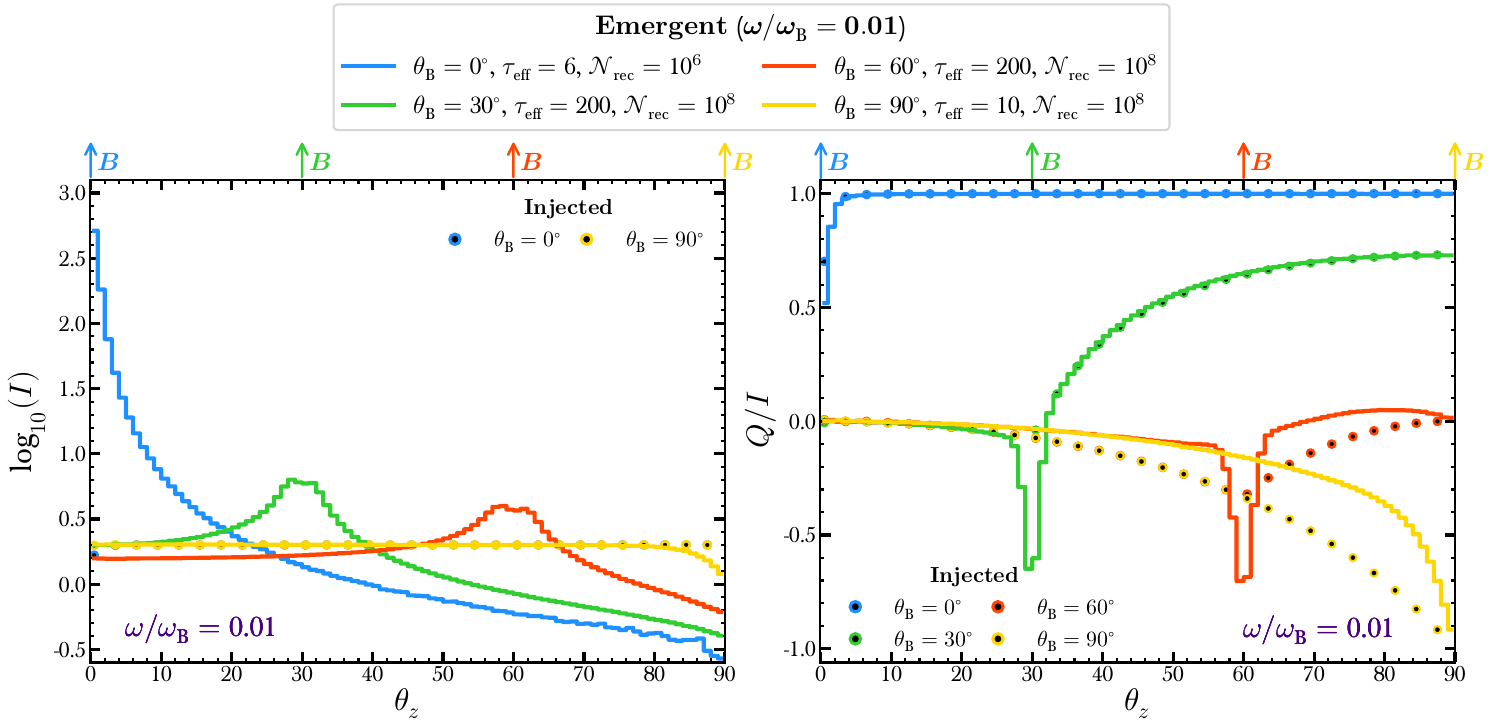}
    \includegraphics[width=1.0\linewidth]{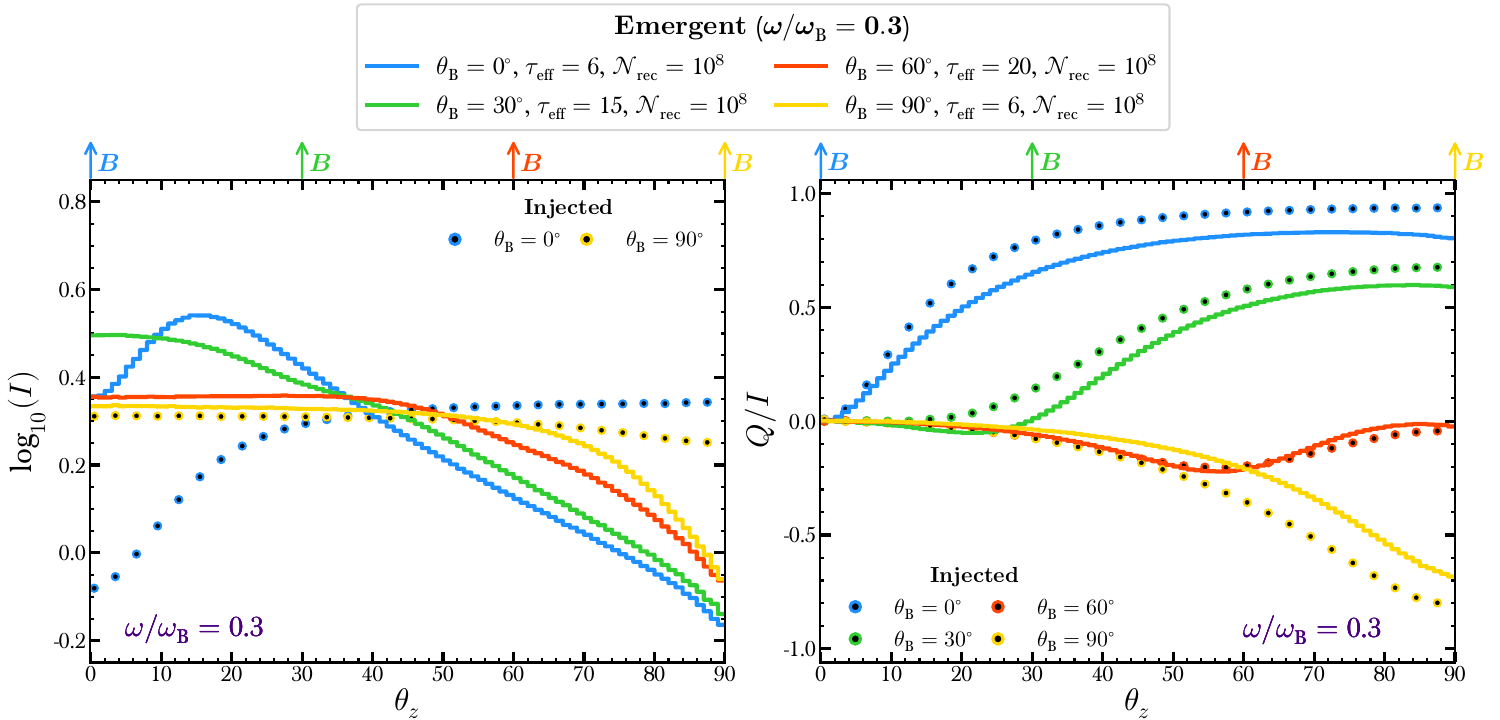}
	\caption{%Azimuth-integrated angular distributions  for logarithmic intensity $\log_{10}{I}$ (left panel) and Stokes parameter $Q/I$ (right panel) as a function of the zenith angle $\theta_z$ for photons emergent at different surface locales corresponding to four different magnetic colatitudes: $\theta_{\rm B} = 0^{\circ}$ (blue), $\theta_{\rm B} = 30^{\circ}$ (green), $\theta_{\rm B} = 60^{\circ}$ (red), and $\theta_{\rm B} = 90^{\circ}$ (orange). Results are obtained at a fixed frequency ratio, $\omega / \wcyc = 0.01$, for different $\mathcal{N}_{\rm rec}$, as labelled.
    Azimuth-integrated angular distributions  for logarithmic intensity $\log_{10}(I)$ (left column) and Stokes parameter $Q/I$ (right column) as a function of the zenith angle $\theta_z$ for photons emergent at different surface locales corresponding to four different magnetic colatitudes: $\theta_{\rm B} = 0^{\circ}$ (blue), $\theta_{\rm B} = 30^{\circ}$ (green), $\theta_{\rm B} = 60^{\circ}$ (red), and $\theta_{\rm B} = 90^{\circ}$ (orange). Results are obtained at fixed frequency ratios, $\omega / \wcyc = 0.01$ (top row) and $\omega/\wcyc = 0.3$ (bottom row), for different $\mathcal{N}_{\rm rec}$, as labelled. The injection distributions are shown with black-centered points that are color-coded as for the emergent histogram distributions.  For \teq{\log_{10}(I)} at left, these injections are displayed only for two magnetic colatitudes, while for \teq{Q/I} (right), they are shown for all colatitudes. The two injection traces for intensity in the top left panel overlap at $\theta_z \gtrsim 1 ^{\circ}$.
    }
	\label{fig:slabs-1d}
\end{figure}

For higher magnetic fields, the radiation emitted at the magnetic pole becomes more collimated as the angular extent of the MSC shrinks.
This is illustrated for $\omega / \wcyc = 0.01$ by the blue curves in the top panels in Figure~\ref{fig:slabs-1d}, which display the logarithmic intensity $\log_{10}(I)$ (left panel) and scaled Stokes parameter $Q/I$ (right panel) as a function of the zenith angle $\theta_z$  at $\theta_{\rm B} = 0 ^{\circ}$. The intensity \teq{I} is normalized using Equation~(7) in \cite{Hu-2022-ApJ}, and peaks at around $\theta_{z} = 1^{\circ}$, as expected for this frequency ratio.  At zenith angles smaller than this, $\parallel$-polarized photons dominate, setting $Q/I > 0$, yet they are still accompanied by some fraction of  $\perp$-polarized photons (i.e., $Q/I < 1$). At $\theta_{z}> 1^{\circ}$, the radiation is fully $\parallel$-polarized, with a linear polarization degree of essentially $100\%$.  These characteristics mirror those in Figure~\ref{fig:fig-slab-2d-theta}.  The distributions in Figure~\ref{fig:slabs-1d} are %also
integrated over the azimuthal angle $\phi_z$, resulting in $U = 0$. Furthermore, as shown in Figure~\ref{fig:fig-slab-2d-theta}, outside the MSC, $V$ is very small for sub-cyclotronic frequencies and therefore is not displayed here. 

To illustrate the variations of anisotropy and polarization across a stellar surface, we also present in Figure~\ref{fig:slabs-1d} the results obtained at three other magnetic colatitudes, corresponding to $\theta_{\rm B} = $ $30^{\circ}$ (green), $60^{\circ}$ (red), and $90^{\circ}$ (orange) for two selected frequency ratios, $\omega/\wcyc = 0.01$ (top row) and $\omega / \wcyc = 0.3$ (bottom row). The number of emergent photons $\mathcal{N}_{\rm rec}$ and the value of the effective optical depth $\tau_{\rm eff}$ at which convergence occurs are indicated for each surface locale.  Along with the emergent distributions (solid curves), the anisotropy and polarization at injection, described in Section~\ref{subsec:magthomscatt}, are  displayed using traces with points. For visual clarity, these injection distributions are only presented at $\theta_{\rm B} = 0^{\circ}$ (pole) and  $\theta_{\rm B} = 90^{\circ}$ (equator) for intensity $I$ (left column), while all four considered locales are plotted for $Q/I$ (right column). For $\omega / \wcyc  = 0.01$ (top left panel), a characteristic signature is the appearance of a peak in the intensity profile around the magnetic field direction, with its prominence decreasing when moving away from the pole. This peak is completely muted at the magnetic equator, mirroring the same behavior evident in the top panels of Figure~\ref{fig:fig-slab-2d-theta-4thB}. The reduction in the intensity peak is caused by two factors: (1) photons are less likely to exit the surface at shallow angles, and (2) the curves all include integrations over the azimuthal angle, with most directions not being near that of the field.  For $\omega/\wcyc = 0.3$ (bottom left panel), the intensity beaming at the pole generates a modest peak at \teq{\theta_z\sim 20^{\circ}}, the approximate angular scale of its MSC.  Moreover, there is essentially no evidence of beaming at other magnetic colatitudes; see also Figure~2 of \cite{Hu-2022-ApJ}. 

Figure~\ref{fig:slabs-1d} also compares  the emergent distributions in \teq{\theta_z} with the injected ones.
At the magnetic pole in the domain where $\omega / \wcyc \ll 1$, the intensity angular distribution of emergent photons exhibits significant deviation from the injected distribution, which is isotropic except within the MSC. This difference arises because the injected distributions were obtained after a fixed total number of scatterings, without accounting for the spatial displacement within the atmospheric slab \citep{Barchas-2021-mnras}. In contrast, at the magnetic equator, where the beaming is diminished by flux weighting and azimuth integration, the emergent configurations resemble those at injection. We note that the two injection traces in the top left panel overlap at $\theta_z \gtrsim 1 ^{\circ}$.

Regarding the polarization in Figure~\ref{fig:slabs-1d}, right, $\parallel$-mode photons that have their electric field vectors in the same plane as the propagation vector and magnetic field vector dominate.
However, the Stokes parameters presented in this Figure and others in this Section are calculated using the  coordinate system defined by  $\hat{\boldsymbol{k}}_{\rm S}$ and $\hat{\boldsymbol{n}}$, i.e., Equation~(\ref{eq:xyz_S}).
Therefore, the Stokes parameter $Q$ varies with both the zenith and azimuth angle (see panel (a) in Figure~\ref{fig:Slab-2d}) depending on the projection of the electric field vector on $\hat{y}_{\rm S}$ and $\hat{x}_{\rm S}$. As a result, the local linear polarization degree $|Q/I|$ is reduced to less than $100\%$ in the mid-latitude and equatorial cases owing to the azimuthal integration.   This property was apparent in a number of figures in \citet{Barchas-2021-mnras} and \citet{Hu-2022-ApJ}.  Furthermore, as discussed in \citep{Barchas-2021-mnras, Baring-ApJ-2025},  in the domain where $\omega \ll \wcyc$, the anisotropy function $\mathcal{A}(\omega)$ asymptotically approaches $-1$, which leads to the injected normalized Stokes parameter $ \hat{Q}_{\omega}(\mu_0)$ in Equation~(\ref{eq:Stokes-inj}) becoming close to 1 for most values of $\mu_0$.  We note that the subscript `0' indicates the injection angle relative to the magnetic field direction, as the injection protocol was designed with respect to $\boldsymbol{B}$, see Equations~(\ref{eq:I_mui}) and~(\ref{eq:Stokes-inj}).  This should be distinguished from the subscript `zi', which refers to the injection angle with respect to the zenith direction, i.e., the vector normal to the atmospheric slab.  This indicates that the majority of photons are injected in the $\parallel$ mode. As a result, the polarization distributions at injection and emergence are largely similar, particularly for $\theta_{\rm B} = 0^{\circ}, 30^{\circ}$, yet with noticeable differences for larger magnetic colatitudes. Notably, the injection traces in the two mid-latitude cases are perfectly aligned with that corresponding to $\theta_{\rm B} = 90^{\circ}$, only diverging when $\theta_z \gtrsim \theta_{\rm B}$; see the top right panel. The similarities between the injected and emergent polarization distributions persist even at higher frequencies, as shown in the bottom right panel for $\omega / \wcyc = 0.3$, although there are obvious modest differences.  Again, the resemblance observed between the prescribed injection, which pertains to the high opacity configuration, and the emergent distributions for both anisotropy and polarization across various cases suggests that the AP injection protocol requires less time to achieve convergence, and hence is generally much more efficient than the IU counterpart; see also \cite{Barchas-2021-mnras, Hu-2022-ApJ, Baring-ApJ-2025}.

\subsection{Polarization Diversity spanning Sub-Cyclotronic and Cyclotronic Domains} \label{subsec:polarization_diversity}

It is instructive to explore the full polarization character at a particular surface locale in greater depth.  To this end, we observe that the
electric field vector at the surface can be decomposed as the superposition of two normal polarization modes (O and X) as 
\begin{equation}
	\boldsymbol{\mathcal{E}}_{\rm S} \; =\; \mathcal{E}_{\rm O, S} \,\hat{e}_{\rm O,S} + \mathcal{E}_{\rm X,S} \,\hat{e}_{\rm X,S} \; =\;  \mathcal{E}_{x, \rm S} \,\hat{x}_{\rm S} + \mathcal{E}_{y,\rm S} \,\hat{y}_{\rm S}\quad ,
 \label{eq:ES}
\end{equation}
in which we identify a polarization coordinate basis particular to each emergent photon:
\begin{equation}
    \hat{e}_{\rm X,S} \; =\; \frac{ \hat{\boldsymbol{B}} \times \hat{\boldsymbol{k}}_{\rm S} }{|\hat{\boldsymbol{B}} \times \hat{\boldsymbol{k}}_{\rm S}|}
    \quad , \quad
    \hat{e}_{\rm O,S} \; =\; \hat{e}_{\rm X,S} \times \hat{\boldsymbol{k}}_{\rm S} 
    \quad .
 \label{eq:eX_eO_surface}
\end{equation}
Clearly, the O mode ($\parallel$) has an electric field vector in the \teq{\hat{\boldsymbol{k}}_{\rm S} - \hat{\boldsymbol{B}}} plane, and the X mode ($\perp$) has its electric field perpendicular to this plane.
In the second identity in Equation~(\ref{eq:ES}), the unit vectors $\hat{x}_{S}$ and $\hat{y}_{\rm S}$ are the slab-normal oriented ones defined in Equation~(\ref{eq:xyz_S}). Let $\psi_{\rm B}$ be the angle between $\hat{e}_{\rm O, S}$ and $\hat{x}_{\rm S}$.  Then one can show using spherical trigonometry that $\psi_{\rm B}$ relates to the magnetic colatitude $\theta_{\rm B}$, photon zenith angle $\theta_{z}$, and photon azimuth angle $\phi_z$ with respect to the  plane of $\boldsymbol{B}$ and $\hat{\boldsymbol{n}}$ via:
\begin{equation}
	\tan \psi_{\rm B} \; =\; \frac{\sin \theta_{\rm B} \sin \phi_z }{ \cos\theta_{\rm B} \sin\theta_{z} -\sin\theta_{\rm B} \cos\theta_{z} \cos \phi_z }
    \quad .
 \label{eq:psiB}
\end{equation}
The Stokes parameters $I$, $Q$, and $U$ can be expressed in terms of the two complex amplitudes $ \mathcal{E}_{\rm O, S}$ and $\mathcal{E}_{\rm X, S} $ that appear in Equation~(\ref{eq:ES}), using the coordinate system defined in Equation~(\ref{eq:xyz_S}):
\begin{eqnarray}
	  I & = &\mathcal{E}_{x,\rm S}\mathcal{E}_{x,\rm S}^{*} + \mathcal{E}_{y, \rm S}\mathcal{E}_{y,\rm S}^{*} = |\mathcal{E}_{\rm O, S}|^2  + |\mathcal{E}_{\rm X, S}|^2 \ , \nonumber  \\[2pt]
 	  Q & = & \mathcal{E}_{x,\rm S}\mathcal{E}_{x,\rm S}^{*} -\mathcal{E}_{y, \rm S}\mathcal{E}_{y,\rm S}^{*} = (|\mathcal{E}_{\rm O, S}|^2  - |\mathcal{E}_{\rm X, S}|^2) \cos2\psi_B - (\mathcal{E}_{\rm O, S}\mathcal{E}^{*}_{\rm X, S} + \mathcal{E}^{*}_{\rm O, S}\mathcal{E}_{\rm X, S}) \sin 2\psi_B \ , \label{eq:IQU_S} \\[2pt]
 	  U & = & \mathcal{E}_{x, \rm S}\mathcal{E}_{y, \rm S}^{*} +\mathcal{E}_{x, \rm S}^{*} \mathcal{E}_{y, \rm S} = (|\mathcal{E}_{\rm O, S}|^2  - |\mathcal{E}_{\rm X, S}|^2) \sin2\psi_B + (\mathcal{E}_{\rm O, S}\mathcal{E}^{*}_{\rm X, S} + \mathcal{E}^{*}_{\rm O, S}\mathcal{E}_{\rm X, S}) \cos 2\psi_B   \ . \nonumber
\end{eqnarray}
Using these forms, the linear polarization degree $\Pi_l$ can then be written as
\begin{equation}
	\Pi_l = \frac{\sqrt{Q^2 + U^2}}{I} =  \frac{\sqrt{ \left( |\mathcal{E}_{\rm O, S}|^2  - |\mathcal{E}_{\rm X, S}|^2 \right)^2 + \left(\mathcal{E}_{\rm O, S}\mathcal{E}^{*}_{\rm X, S} + \mathcal{E}^{*}_{\rm O, S}\mathcal{E}_{\rm X, S}\right)^2}}{|\mathcal{E}_{\rm O, S}|^2  + |\mathcal{E}_{\rm X, S}|^2}\ . \label{eq:pol_deg_S}
\end{equation}	
In the $\omega \ll \wcyc$ domain, after scatterings, $|\mathcal{E}_{X,\rm S}| \approx  0$ for photons emerging from the surface, and the scaled Stokes parameters $Q/I$ and $U/I$ take the form of simple trigonometric functions:
\begin{equation}
	\left(\frac{Q}{I}\right)_{\omega \ll \wcyc} \approx \cos2\psi_{\rm B}  
    \quad , \quad  
    \left(\frac{U}{I}\right)_{\omega \ll \wcyc} \approx \sin2\psi_{\rm B}  \quad , 
 \label{eq:QandU}
\end{equation}
so that the linear polarization degree is 
\teq{\Pi_l \approx 1}.  An alternative measure of polarization is the \teq{OX}-mode polarization fraction
\begin{equation}
    f_{\rm OX} \;\equiv\; \frac{|\mathcal{E}_{\rm O,S}|-|\mathcal{E}_{\rm X,S}|}{|\mathcal{E}_{\rm O,S}|+|\mathcal{E}_{\rm X,S}|} \quad ,
 \label{eq{OX_polfrac}}
\end{equation}
which provides a clear measure of which polarization eigenmode dominates the photon population emergent from the slab.  In general, \teq{f_{\rm OX}} differs in value from \teq{\Pi_l}, yet is also numerically close to unity when $\omega \ll \wcyc$.

\begin{figure}[t]
	\centering
    \includegraphics[width=1.0\linewidth]{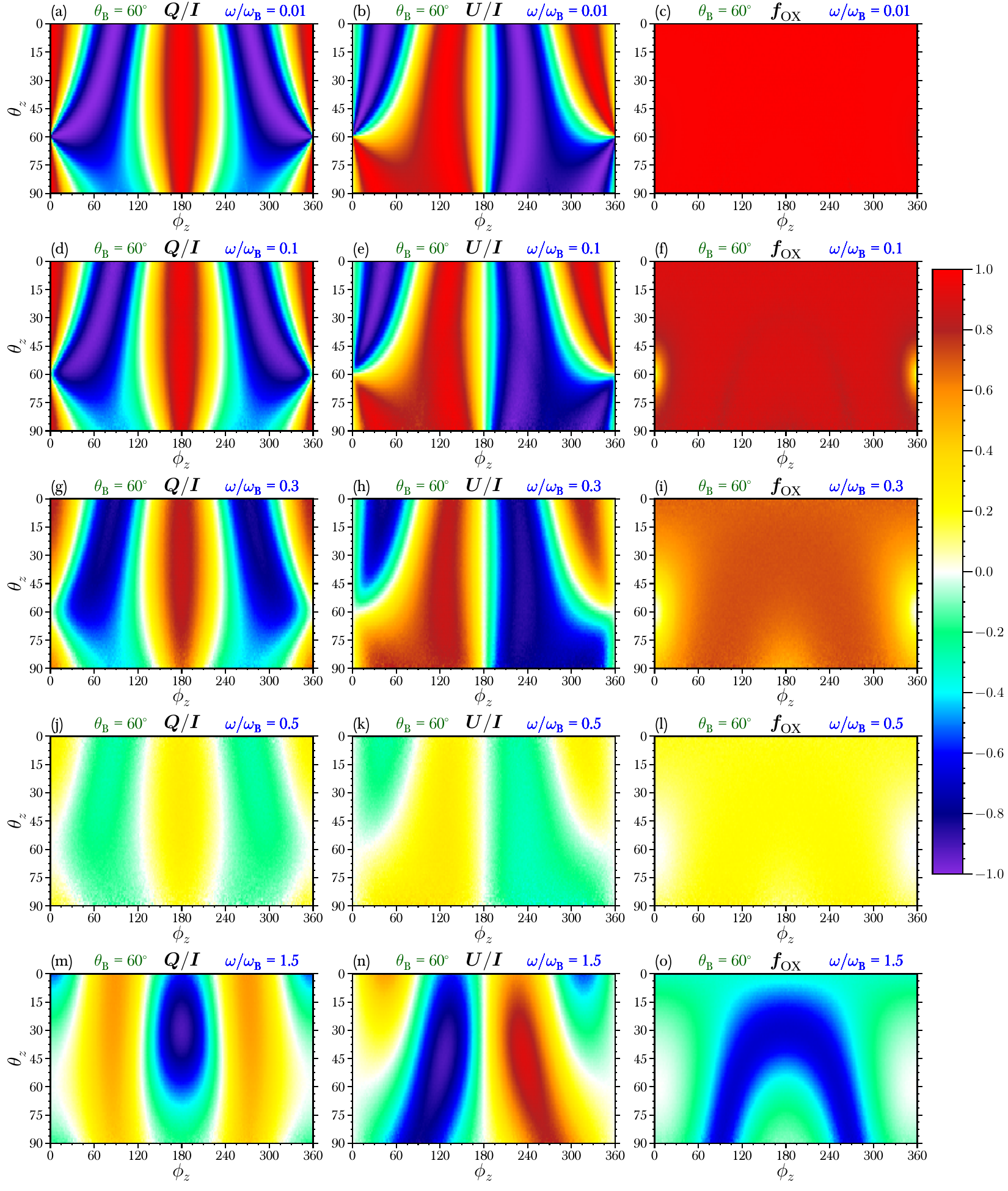}
	\caption{Two-dimensional angular distributions as a function of zenith and azimuthal angles, $\theta_z$ and $\phi_z$, of Stokes parameters $Q/I$ (left column), $U/I$ (middle column), and $f_{\rm OX} = \{ |\mathcal{E}_{\rm O,S}|-|\mathcal{E}_{\rm X,S}| \}/ \{|\mathcal{E}_{\rm O,S}|+|\mathcal{E}_{\rm X,S}|\}$ (right column) for $\omega/\wcyc = 0.01$ ($\tau_{\rm eff} = 200$, panels a-c), $\omega / \wcyc = 0.1$ ($\tau_{\rm eff} = 30$, panels d-f), $\omega / \wcyc = 0.3$ ($\tau_{\rm eff} = 20$, panels g-i), $\omega / \wcyc = 0.5$ ($\tau_{\rm eff} = 10$, panels j-l),  and $\omega / \wcyc = 1.5$ ($\tau_{\rm eff} =6$, panels m-o)  at $\theta_{\rm B} = 60^{\circ}$ for $\mathcal{N}_{\rm rec} = 10^8$ recorded photons.
    }
	\label{fig:Slab-2d}
\end{figure}

Some of these polarization measures are illustrated in Figure~\ref{fig:Slab-2d}.  In the top row, the two-dimensional distributions of $Q/I$ (panel a), $U/I$ (panel b), and the $OX$-mode polarization fraction $f_{\rm OX}$ (panel c) are presented as functions
of the emergent direction angles $\theta_z$ and $\phi_z$.  These sky maps are for $\omega /\wcyc = 0.01$ and  $\mathcal{N}_{\rm rec} = 10^8$ photons emerging from the upper slab surface, and we chose $\theta_{\rm B} = 60^{\circ}$ for this illustrative example.  The ratio $f_{\rm OX}$ is always close to one, indicating that $\mathcal{E}_{\rm X,S} \approx 0$ and the emergent radiation is dominated by the O-mode photons, as one expects from the magnetic Thomson differential cross section. Moreover, the Stokes parameters $Q/I$ in panel (a) and $U/I$ in panel (b) indeed behave as $\cos 2 \psi_{\rm B}$ and $\sin 2 \psi_{\rm B}$, respectively, as expressed in Equation~(\ref{eq:QandU}), with $\psi_{\rm B}$ being given in Equation~(\ref{eq:psiB}).  One can also notice that, for each value of $\theta_z$, $\psi_{\rm B}$ in Equation~(\ref{eq:psiB}) is anti-symmetric about $\phi_z = 0$. This explains the  symmetric   pattern in  $Q/I \approx \cos 2 \psi_{\rm B}$ (anti-symmetric in $U/I \approx \sin 2 \psi_{\rm B}$) observed in  panel (a)  and panel (b) of Figure~\ref{fig:Slab-2d}.
Additionally, the values of $\theta_z$ and $\phi_z$ at which $Q$ ($U$) equals zero correspond to the solutions of Equation~(\ref{eq:psiB}) where $\psi_{\rm B} = \pi (2n +1)/4$ %(and $\psi_{\rm B} = \pi (2n +1)/2$)
(and {$\psi_{\rm B} = \pi n/2$}), with $n \in \mathbb{Z}$.
Finally, note that the dominance of $\parallel$-photons always applies in this highly magnetic domain regardless of the magnetic colatitude of the atmospheric slab. Therefore, $f_{\rm OX} \approx 1$, and the Stokes parameters, $Q/I$ and $U/I$, behave according to Equation~(\ref{eq:QandU}) for any $\theta_{\rm B}$.
 
For comparison,  we present in the next three rows in Figure~\ref{fig:Slab-2d} the distributions for three higher sub-cyclotronic frequency ratios: $\omega/ \wcyc = 0.1$ (panels d-f), $\omega/ \wcyc = 0.3$ (panels g-i), and $\omega / \wcyc = 0.5$ (panels j-l). For $\omega/\wcyc = 0.1$, except for directions around the magnetic field, i.e., $\theta_z = \theta_{\rm B} = 60^{\circ}$ and $\phi_z = 0^{\circ}, 360^{\circ}$, the fraction $f_{\rm OX}$ is aways greater than  $\sim 0.8$ (panel f), indicating the strong dominance of O-mode photons. As a result, the patterns of $Q/ I$ (panel d) and $U/I$ (panel e) are essentially identical to those obtained in the case where $\omega / \wcyc = 0.01$ (panels a-c). As the frequency ratio further increases, the  degree of linear polarization decreases. For,  $\omega /\wcyc = 0.3$ (panels g-i), $f_{\rm OX}$ is less than about 0.6 and then drops below 0.2 for $\omega /\wcyc = 0.5$ (panels j-l).  This trend is caused by the progressive reduction of the disparity between the scattering cross sections for the two linear polarization modes as the cyclotron frequency is approached.
 
Finally, in the bottom row of Figure~\ref{fig:Slab-2d}, i.e., panels (m)-(o), we display results for a super-cyclotronic case,  $\omega /\wcyc = 1.5$, again for 10$^8$ emergent photons.  This frequency ratio is germane to neutron stars of low magnetization, such as X-ray dim isolated neutron stars (XDINS). As discussed in \citet{Barchas-2021-mnras}, the polarization characteristics are inverted in sign as the frequency ratio goes from $\omega/\wcyc < 1/\sqrt{3}$ to $\omega/\wcyc > 1/\sqrt{3}$.   As a result,  the emergent radiation for $\omega /\wcyc = 1.5$ is governed by $\perp$-photons ($f_{\rm OX} < 0$), as shown in panel (o). Moreover, $Q/I$ (panel m) and $U/I$ (panel n) respectively exhibit somewhat similar patterns as those in panels (a) and (b), except that they possess the opposite signs.  Nevertheless, the magnitudes of $Q$ and $U$ at $\omega/\wcyc = 1.5$ are noticeably lower than those obtained for magnetars. This is because in the case of low magnetizations, although $\perp$-photons dominate,  the contribution from $\parallel$-photons in the radiation configuration is considerable.  These contrasting polarization characteristics between magnetars and XDINS should be prevalent observationally, although since XDINS are inherently much fainter, it will require next-generation X-ray polarimeters to probe their Stokes parameter signatures.

\begin{figure}[h]
    \centering
    \includegraphics[width=\linewidth]{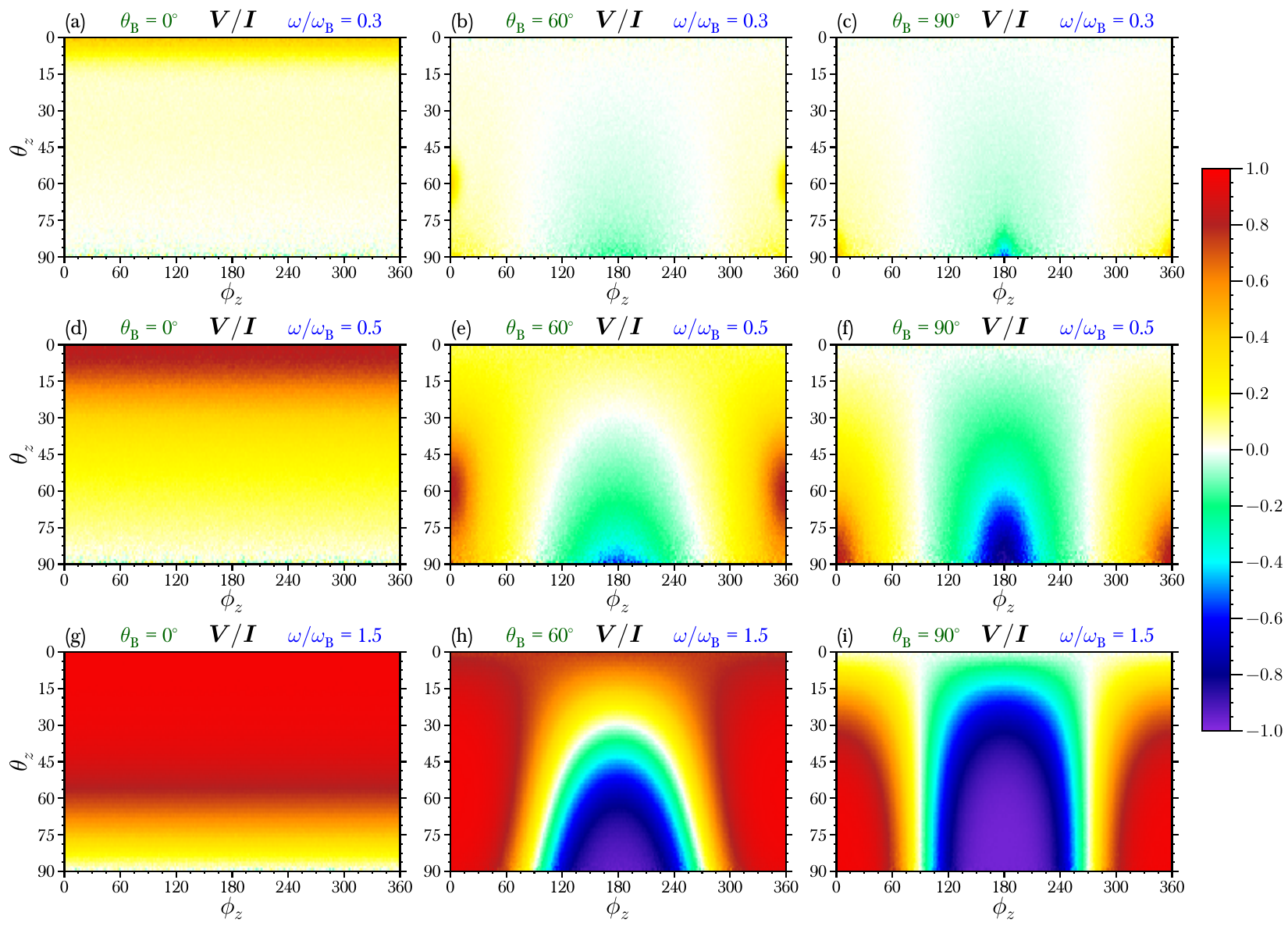}
    \caption{Two-dimensional angular distributions as a function of zenith and azimuthal angles, $\theta_z$ and $\phi_z$, of the Stokes parameter $V/I$ for three selected frequency ratios: $\omega / \wcyc = 0.3$ (top row), $\omega / \wcyc =0.5$ (middle row), and $\omega / \wcyc = 1.5$ (bottom row). For each $\omega / \wcyc$, three magnetic colatitudes, corresponding to $\theta_{\rm B} = 0^{\circ}$ (left column), $\theta_{\rm B} = 60^{\circ}$ (middle column), and $\theta_{\rm B} = 90^{\circ}$ (right column), are considered. The effective optical depth $\tau_{\rm eff}$ values for the runs at the pole ($\theta_{\rm B} = 0^{\circ}$) and equator ($\theta_{\rm B} = 90^{\circ}$) are fixed at $\tau_{\rm eff} = 6$ for all frequencies, while $\tau_{\rm eff}$ values for different $\omega / \wcyc$ in the mid-latitude case ($\theta_{\rm B} = 60^{\circ}$) are as indicated in Figure~\ref{fig:Slab-2d}. See also Table~3 in \citet{Baring-ApJ-2025} for the $\tau_{\rm eff}$ values. }
    \label{fig:Slab-2d-V}
\end{figure}

The complex electric vector formalism used in the {\sl MAGTHOMSCATT} simulation offers another distinctive advantage by allowing for identifying the intricate interplay between linear and circular polarizations, an aspect often overlooked in many works on radiative transfer.	Since  circular polarization is negligible in the $\omega /\wcyc \ll 1$ domain (see \citet{Hu-2022-ApJ} and \cite{Baring-ApJ-2025}),  in Figure~\ref{fig:Slab-2d-V}, we show the distributions of $V/I$ only for $\omega / \wcyc > 0.1$ domains, specifically, $\omega/\wcyc = 0.3$ (top row), $\omega / \wcyc = 0.5$ (middle row), and $\omega / \wcyc = 1.5$ (bottom row). The results obtained with $\theta_{\rm B} = 60^{\circ}$ (middle column) are compared with those obtained at the magnetic pole (left column) and the magnetic equator (right column). In all cases, it is clear that $|V|$ is maximized along the $\boldsymbol{B}$ direction -- with $V$ being positive (negative) if the photon propagation vector $\boldsymbol{k}_{\rm S}$ is parallel (anti-parallel) to $\boldsymbol{B}$ -- as a consequence of electronic gyrational motion in a magnetic field.  At the magnetic equator, $V$ is anti-symmetric around $\phi_z = 90^{\circ}$ and $\phi_z = 270^{\circ}$. Therefore, the azimuthally-averaged $V$ is always zero at the equator; see also \citet{Hu-2022-ApJ}. At $\omega  / \wcyc = 0.3$, circular polarization is small since the photon frequency is still far from the cyclotron resonance. As the photon/cyclotron frequency ratio approaches unity, the circular polarization becomes more significant.  Indeed, at $\omega / \wcyc = 0.5$, which is near the bifurcation point ($\omega /\wcyc = 1/ \sqrt{3}$), where the cross sections for linear and circular polarizations coalesce \citep{Barchas-2021-mnras, Baring-ApJ-2025}, circular polarization becomes significant with $|V / I| \lesssim 0.8$. At $\omega / \wcyc = 1.5$, $|V / I | \lesssim 1$. Furthermore, the interplay between circular and linear polarization is evident when comparing Figures~\ref{fig:Slab-2d-V}(e) and (h) with Figures~\ref{fig:Slab-2d}(l) and (o).  In particular, at $\omega / \wcyc = 1.5$, there is an evident anti-correlation between $|V/I|$ (panel h, Figure~\ref{fig:Slab-2d-V}) and $|f_{\rm OX}|$ (panel o, Figure~\ref{fig:Slab-2d}), where small $|V/I|$ corresponds to large $|f_{\rm OX}|$ and vice versa. At  $\omega / \wcyc = 0.5$, this correlation is still present but much less apparent; see Figures~\ref{fig:Slab-2d}(l) and \ref{fig:Slab-2d-V}(e).

To summarize, the results in this section demonstrate that  radiation in the $\omega / \wcyc \ll 1$ domain is very highly polarized. In particular, the dominance of $\parallel \rightarrow \parallel$ component in the magnetic Thomson scattering differential cross section results in the prevalence of photons in the $\parallel$ polarization mode, except within a small angle, $\sim\theta_{\rm MSC} = \omega/\wcyc $, around the magnetic field direction.
For X-ray emission from magnetars, whose surface magnetic field strength is typically around $10^{14} - 10^{15}$ G, the frequency ratio $\omega / \wcyc$  is of the order of  $10^{-3}-10^{-4}$. Therefore, one can expect that the radiation configuration emerging from an atmospheric surface will be almost entirely in the $\parallel$ mode. 
These properties are well characterized by adequately low frequency ratio values, $\omega / \wcyc \sim 10^{-2}$, where  $\theta_{\rm MSC} \sim \omega /\wcyc$ is sufficiently small that the contribution of photons inside the magnetic scattering cone is negligible when generating pulse profiles of intensity and polarization, hence circumventing the need to address slower runs at lower frequency ratios $\omega/\wcyc$; this element is addressed towards the end of Section~\ref{sec:new_protocol}.

\subsection{The non-magnetic $\omega / \wcyc \gg 1 $ domain}
 \label{subsec:weak-field}

\begin{figure}[t]
    \centering
    \includegraphics[width=\linewidth]{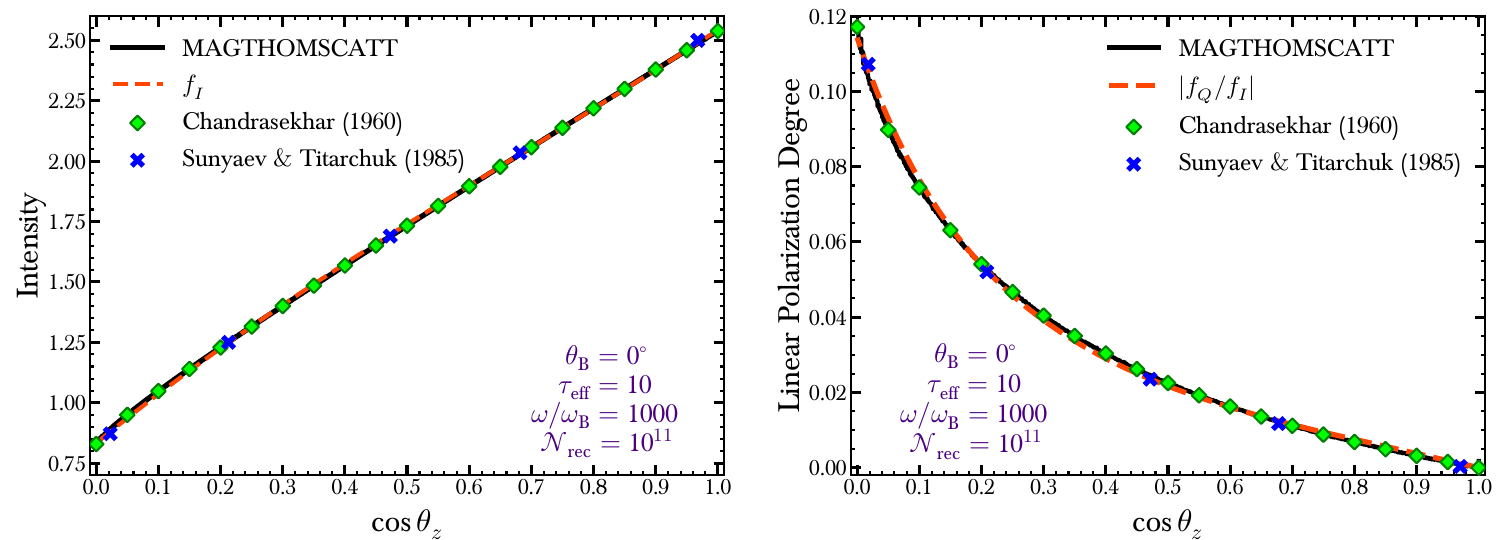}
    \caption{Angular distributions for intensity (left panel) and linear polarization degree, $\Pi_l = \sqrt{Q^2 + U^2}/I = \left|Q/I\right|$ (right panel), as functions $\cos\theta_z$ from the {\sl MAGTHOMSCATT} simulation (black histograms). Note that  $U = 0$ due to the integration over azimuthal angles around the zenith. The simulation was performed for $\omega / \wcyc = 1000$, $\tau_{\rm eff} = 10$, $\mathcal{N}_{\rm rec} = 10^{11}$, and $\theta_{\rm B} = 0^{\circ}$. The distributions are binned at a $\theta_z$ resolution of $0.0625^{\circ}$. The red dashed lines present the analytical fit functions in Equation~(\ref{eq:fit}), while the traces with green diamonds and blue crosses respectively correspond to results from  \cite{Chandrasekhar-1960-book} and \cite{Sunyaev-1985-AandA} for non-magnetic Thomson transport. The intensity (left panel) in \cite{Chandrasekhar-1960-book} (green diamonds) and \cite{Sunyaev-1985-AandA} (blue crosses) are scaled by a factor of 2 and 2.54, respectively.}
    \label{fig:wr1000_I_Q}
\end{figure}

The anisotropy and polarization properties in the super-cyclotronic domain were previously studied in \cite{Barchas-2021-mnras} and \cite{Hu-2022-ApJ} at $\omega / \wcyc = 10$ and $\omega / \wcyc = 3$, respectively. In this paper, we present similar findings but at a much higher frequency ratio, $\omega / \wcyc = 1000$, along with improved statistics, $\mathcal{N}_{\rm rec} = 10^{11}$. This frequency ratio represents the non-magnetic $\omega / \wcyc \gg 1$ regime, where the magnetic Thomson scattering cross section approaches the non-magnetic value, $\sigma \rightarrow \sigma_{\rm T} = 8 \pi r_0^2  / 3$, with $r_0$ being the classical electron radius.  One can expect the magnetic field direction in this context to have negligible impacts on the scattering transport. For this reason, we show in Figure~\ref{fig:wr1000_I_Q} only the angular distributions for the case of $\theta_{\rm B} = 0^{\circ}$, where the magnetic field vector is along the zenith direction, noting that all other $\theta_{\rm B}$ values generate essentially the same results for $\omega / \wcyc \gg 1$.   The circular polarization degree in this case is negligible, $|V/ I| < 4 \times 10^{-3}$, and therefore, is not shown.

The left panel of Figure~\ref{fig:wr1000_I_Q} presents the angular distributions for intensity as functions of the cosine of the emerging zenith angle, $\cos \theta_{z}$.  As remarked in \cite{Barchas-2021-mnras} and \cite{Hu-2022-ApJ},  the intensity of radiation in this scenario exhibits significantly less anisotropy compared to the sub-cyclotronic case. The intensity is maximum along the zenith direction, where $I \approx 2.54$, and decreases to $I \approx 0.82$ along the surface. On the other hand, the linear polarization degree (right panel), $\Pi_l = \sqrt{Q^2 + U^2} / I = \left|Q/I\right|$, is maximum when $\cos \theta_z = 0$, where $\Pi_l \approx 11.6\% $,   and monotonically decreases to 0 when $\cos \theta_z = 1$. We note that $U = 0$ due to the integration over the azimuthal angles around the zenith for our chosen coordinate system.  We used an angular bin size of $\Delta \theta_{z} = 0.0625^{\circ}$  to accurately capture the behavior of the linear polarization degree in the region near $\cos\theta_z = 0$ (i.e., the slab horizon).
 
We compare our results (black histograms)  with the numerical solutions to the non-magnetic radiative transfer integro-differential equation (RTE) presented in Table XXIV on page 248 of the textbook by \cite{Chandrasekhar-1960-book} (traces with green diamonds). For the intensity, we multiply the result by \cite{Chandrasekhar-1960-book} by a constant factor of 2. Moreover, in Figure~\ref{fig:wr1000_I_Q}, we also display the results by \citet{Sunyaev-1985-AandA}, which were obtained by solving the RTE for the non-magnetic Thomson domain in the context of accretion discs near black holes. In particular, the traces with blue crosses in Figure~\ref{fig:wr1000_I_Q} are taken from Figures~4 and 5 in \cite{Sunyaev-1985-AandA} in the case of infinite optical depth, multiplying the intensities of \cite{Sunyaev-1985-AandA} by a constant factor of 2.54. The agreement between our results with these historical benchmarks are excellent for both the intensity and polarization distributions.  Comparing also to the results corresponding to $\omega / \wcyc = 10$ in \cite{Barchas-2021-mnras} (see their Figure~3), where $\Pi_l \approx 15 \%$ at $\theta_z = 90^{\circ}$,  we found that the linear polarization degree obtained using $\omega / \wcyc = 1000$ agrees more closely to the non-magnetic Thomson transport results by \cite{Chandrasekhar-1960-book, Sunyaev-1985-AandA}, as one may expect. The remarkable consistency between our results and those of \cite{Chandrasekhar-1960-book, Sunyaev-1985-AandA}, together with the agreement observed with the results from \cite{Whitney-1991-ApJS} at different frequency ratios reported in \cite{Barchas-2021-mnras}, are robust validations for our {\sl MAGTHOMSCATT} simulation.

The non-magnetic domain addressed in this Section is relevant to X-ray emission from surfaces of  millisecond pulsars (MSPs) and highly-magnetic white dwarfs. For potential applications to numerical simulations and data interpretation, such as pulse profile modeling for MSPs, we also provide the following fitting functions for the Stokes \teq{I} and \teq{Q} parameters:
\begin{equation}     
   f_I(x) \; = \; \frac{1}{7} \left(\frac{35}{6} + \frac{44}{3}x - \frac{26}{5}x^2 + \frac{5}{2}x^3 \right) \quad , \quad
   f_Q(x) \; = \; -\frac{1}{63}\left(6 - 5 x + 5x^2\right)(1 - x) \quad , % new fit
    \label{eq:fit}
\end{equation}
where the intensity and linear polarization degree are respectively given by $I = f_I(\cos\theta_z)$ and $\Pi_l = |f_Q(\cos\theta_z) / f_I(\cos\theta_z)|$. These analytical fitting functions are shown by the red dashed lines in Figure~\ref{fig:wr1000_I_Q},  and we can see that the precisions of the fits are both very good. 

\section{Computational Efficiency in the Magnetar Domain}
%\section{Approximations to High-opacity Configurations in the Magnetar Domain}
\label{sec:new_protocol}

It was shown that AP injection protocol described in Section~\ref{subsec:magthomscatt}, with the empirical expressions for coefficients $\mathcal{A}(\omega)$ and $\mathcal{C}(\omega)$ being provided in \citet{Baring-ApJ-2025}, significantly enhances the simulation efficiency, notably when $\omega / \wcyc \ll 1$, compared to the  IU counterpart; see Table~3 in \citet{Baring-ApJ-2025}.  Nevertheless, in this frequency ratio range, which is germane to soft X-ray emission from magnetar surfaces, the significant disparity of the cross sections for different polarization modes and propagation directions requires much higher Thomson optical depths $\taut$ for convergence; see the discussions in Section~\ref{subsec:magnetar}. This is especially true in surface regions where \(\theta_{\rm B} < 90^{\circ}\). Consequently, despite the improvements achieved with the AP injection protocol, the simulation remains relatively inefficient in this specific domain.
 
\subsection{A refined injection protocol}
 \label{subsec:APstar}

The results presented in Section~\ref{sec:emergent} demonstrate that the radiation is nearly fully polarized in the $\parallel$-mode outside of the MSC at convergence. This is consistent with the analysis in \citet{Baring-ApJ-2025}. Within the MSC,  the significant reduction in the cross sections, regardless of  photon polarization modes, leads to a pronounced beaming effect along the magnetic field direction; see Figures~\ref{fig:fig-slab-2d-theta}-\ref{fig:slabs-1d}. 
As discussed earlier, in the considered frequency domain, $\sigma_{\perp} \ll  \sigma_{\parallel}$, except for the direction along $\boldsymbol{B} $.
As a consequence, if the optical depth is not sufficiently large, the emergent polarization information at  $\arccos{(\hat{\boldsymbol{k}}_{\rm S}\cdot \hat{\boldsymbol{B}})} > \thetaMSC$ is contaminated by un-scattered $\perp$-mode photons from the injection, as these photons can free stream out without scattering.  To eliminate these un-scattered $\perp$-mode photons and achieve the expected emergent polarization distribution, one straightforward approach is to increase the effective optical depth in the simulation. Alternatively, a condition can be imposed in the simulation such that only photons scattering at least once are taken into account, i.e., $n_{\rm scat, min} = 1$, as was done for the results obtained in Section~\ref{sec:emergent}.

The latter approach is efficient for attaining the desired converged polarization configuration.  However,  the requirement that each photon must scatter at least once causes photons injected at small angles to deflect from the magnetic field direction. Consequently, a significant number of scatterings are still necessary to acquire the peak expected along the magnetic field for the convergent anisotropy; see Figures~\ref{fig:fig-slab-2d-theta} and \ref{fig:fig-slab-2d-theta-4thB}. For example, in the case of $\omega / \wcyc = 0.03$ and $\theta_{\rm B} = 0^{\circ}$ shown in Figure~\ref{fig:fig-slab-2d-theta}, the average number of scatterings per photon is $1.6 \times 10^6$, with $\tau_{\rm eff} = 6$. Nevertheless,  the spatial distributions displayed in Figure~\ref{fig:distribution} indicates that the Markovian domain is reached after $\sim 5 \times 10^5$ scatterings. Therefore, we can expect that, if only scattered photons are considered, then both the convergent anisotropy and polarization distributions outside the the MSC can be achieved at a lower effective optical depth than that used in Figure~\ref{fig:fig-slab-2d-theta}. On the other hand, the behaviors within the MSC, particularly the beaming effect, can be achieved at a smaller $\tau_{\rm eff}$ if we account for photons that emerge un-scattered within this region.

Motivated by these characteristics, to explore possible enhancements of code efficiency, we incorporated into the simulation a new protocol, called AP*, where we combined the AP injection protocol outlined in Section~\ref{subsec:magthomscatt} with the following prescription for the minimum scattering number:
\begin{equation}
	n_{\rm scat, min } \quad = \quad
		\begin{cases} 
			5H(\theta_i -2\thetaMSC) & \text{if $\theta_{\rm B} \leq 2^{\circ}$}  \\
		   	5H(\theta_i -0.1) & \text{if $\theta_{\rm B} > 2^{\circ}$}
		\end{cases} \quad ,
	\label{eq:nscat_min}
\end{equation}
where $\theta_i$ is the photon incident angle (in radians) with respect to the magnetic field direction, and $H(x)$ is the Heaviside step function, with $H(x) = 1$ if $x\geq 0$, and zero otherwise. From Equation~(\ref{eq:nscat_min}), it is evident that un-scattered photons are taken into account only if they travel within small angles relative to the direction of the magnetic field. To enhance simulation speed and reduce the effective optical depth for convergence, we have set $n_{\rm scat, min}$ to 5, rather than 1, for photons propagating at larger angles to the magnetic field. This adjustment aims to more effectively eliminate $\perp$-mode photons in these directions, simulating the same effects as having a large optical depth. We note that for surface locales away from the magnetic pole, the intensity peak becomes broadened since most directions do not align closely with the magnetic field vector. Thus, for these cases, the condition $n_{\rm scat} \geq 5$ is only applied for photons traveling at an angle of $\theta_{i} > 0.1$ with respect to $\boldsymbol{B}$. We explored other values for $n_{\rm scat, min }$ and the dividing value for $\theta_{\rm B}$ in Equation~(\ref{eq:nscat_min}), and found that the above choice was optimal.

\begin{figure}[t]
 	\centering
 	\includegraphics[width=1\linewidth]{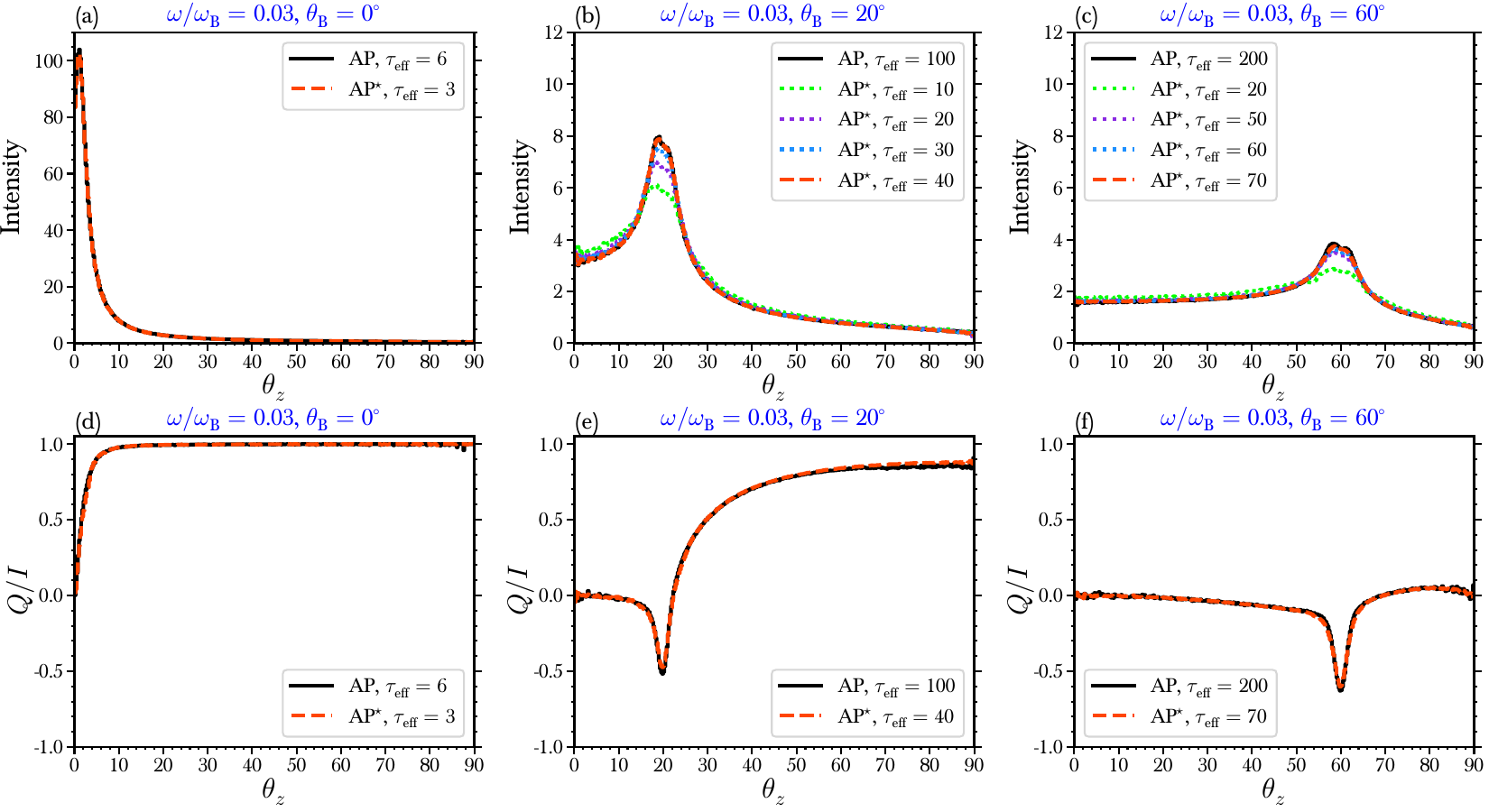}
 	\caption{Angular distributions of intensity (top row) and $Q/I$ (bottom row) obtained with the standard AP protocol (black solid lines), where $n_{\rm scat, min} = 1$ for all $\theta_i$ and $\theta_{\rm B}$, in comparison with those obtained with the AP* protocol, where $n_{\rm scat, min}$ is given by Equation~(\ref{eq:nscat_min}), for $\omega / \wcyc = 0.03$. The results are displayed for three selected $\theta_{\rm B}$ values: $\theta_{\rm B} = 0^{\circ}$ (left column), $\theta_{\rm B} = 20^{\circ}$ (middle column), and $\theta_{\rm B} = 60^{\circ}$ (right column). The results at convergence in the AP* protocol are shown in red dashed lines. In panels (b) and (c), we also present for the AP* protocol distributions at three $\tau_{\rm eff}$ values smaller than the convergence one.}
 	\label{fig:AP_approach_0.03}
\end{figure}

The comparisons between the emergent distributions for intensity (top row) and $Q/I$ (bottom row) obtained using the standard AP protocol, where $n_{\rm scat, min} =1$  for all $\theta_i$ and $\theta_{\rm B}$, with those in the AP* protocol, where $n_{\rm scat, min}$ is given in Equation~(\ref{eq:nscat_min}), are illustrated in Figure~\ref{fig:AP_approach_0.03} for $\omega / \wcyc = 0.03$. Three different magnetic colatitudes, corresponding to $\theta_{\rm B} = 0^{\circ}$ (left column), $\theta_{\rm B} = 20^{\circ}$ (middle column), and $\theta_{\rm B} = 60^{\circ}$ (right column), are considered. The black solid lines (red dashed lines) show the convergent configurations for intensity and $Q/I$ obtained in the AP (AP*) protocol with an effective optical depth of respectively 6 (3), 100 (40), and 200 (70) for the three $\theta_{\rm B}$ values.  Again, $V/I \approx 0$ in this domain, and $U / I = 0$ due to the integration around the zenith direction. It is evident that convergence is established faster for the AP* protocol than in the standard AP counterpart. In panels (b) and (c) of Figure~\ref{fig:AP_approach_0.03}, we also display the AP* intensity distributions for three lower $\tau_{\rm eff}$ values than the required one for convergence, as shown by dotted lines in green, purple, and blue. We can see that, near the convergence, the variation in the distributions when increasing $\tau_{\rm eff}$ is minimal. For visual clarity, this comparison is not included for all panels.

We extended the above comparison to other sub-cyclotronic frequencies and magnetic colatitudes. Table~\ref{tab:tau_eff} summarizes the values of effective optical depths at which the convergence configuration is achieved, denoted as $\tau_{\rm eff, AP}$ and $\tau_{\rm eff, AP*}$, as well as the associated run time ratio, $t_{\rm AP} / t_{\rm AP*}$, for four frequency ratios, ranging between $0.01 \leq \omega / \wcyc \leq 0.3$, and eleven $\theta_{\rm B}$ values, ranging from the magnetic pole to the equator. The AP* protocol is shown to accelerate the speed of the simulation up to a factor of $\sim 9$ for the frequency ratios considered. In particular, the enhancement is significant when $\omega / \wcyc < 0.1$ and $\theta_{\rm B} < 90^{\circ}$. In the cases of higher frequency ratios and $\theta_{\rm B} = 90^{\circ}$, the AP* provides no material improvement over AP. This behavior aligns with our expectation because the distributions at injection are close to those at emergence  for the latter cases, as shown in Figure~\ref{fig:slabs-1d}, and therefore, the simple $n_{\rm scat, min} =1$ condition is sufficient.  We also remark that $\theta_{\rm B} \gtrsim 60^{\circ}$ locales are generally less likely to be sampled in neutron stars whose surface emission predominantly arises in proximity to the poles.

\begin{deluxetable}{lr||c c c c c c c c c c c}[t]
\label{tab:tau_eff}
\tablecaption{Effective optical depths and ratios of run times in AP and AP* protocols }
%\tabletypesize{\scriptsize}
\tablewidth{0pt} 
\tablehead{
\colhead{} &\colhead{$\theta_{\rm B} =$} & \colhead{$0^{\circ}$} & \colhead{$2^{\circ}$} &
\colhead{$5^{\circ}$} & \colhead{$10^{\circ}$} & \colhead{$20^{\circ}$}
& \colhead{$30^{\circ}$} & \colhead{$45^{\circ}$}
& \colhead{$60^{\circ}$}
& \colhead{$70^{\circ}$}
& \colhead{$80^{\circ}$}
& \colhead{$90^{\circ}$}
}
\startdata
{  } & $\tau_{\rm eff, AP}$ & 6  &8   &8  &8  &10  &15  &15  & 20& 15&12 & 6 \\
{  $\omega / \wcyc = 0.3$} & $\tau_{\rm eff, AP*}$ & 6 &  6 & 8 & 8 &10  &15  &15  &20 & 12&12 & 6\\
{  }& $t_{\rm AP} / t_{\rm AP*}$&1.6   &1.8  &1.7 & 1.6 &0.9  &0.9  &1.0  & 0.9&1.3 &0.9 &  1.5\\ 
\hline
{  } & $\tau_{\rm eff, AP}$ &  6& 6 & 6& 8& 20 &30 &35  & 30& 30&20 & 10\\
{  $\omega / \wcyc = 0.1$} & $\tau_{\rm eff, AP*}$ &4& 4 &4 &6 & 15 &20 &  30&25 &25 &20 &10  \\
{  }& $t_{\rm AP} / t_{\rm AP*}$ &2.2 & 2.3& 2.1&2.6 &1.7&2.1 &1.3 &2.4  &1.4 &1.0 &  1.0 \\ 
\hline
{  } & $\tau_{\rm eff, AP}$ &  6&6 & 12& 30 & 100 &150 & 200  & 200&150 & 150&10 \\
{  $\omega / \wcyc = 0.03$}& $\tau_{\rm eff, AP*}$ & 3 & 3 & 6  & 12 & 40 &  60&  70 & 70 & 60 & 60 & 10\\
{  }& $t_{\rm AP} / t_{\rm AP*}$ &3.9 &3.7  &3.9 & 5.4 & 5.4& 5.4& 6.8  &7.2 & 8.7& 5.7&  0.6 \\ 
\hline
{  } & $\tau_{\rm eff, AP}$  & 6  & 8 &  40& 110&  200&200 & 200 & 200& 200& 200& 10 \\
{  $\omega / \wcyc = 0.01$}& $\tau_{\rm eff, AP*}$ & 3  & 4 & 15 &40 & 70 & 70 & 70  &70 & 70&70 & 10\\
{  }& $t_{\rm AP} / t_{\rm AP*}$&4.4 & 2.4&6.2 & 6.5 &7.4 & 7.0 & 7.2& 7.3&7.0 &7.8 &  0.6
\enddata
%\vspace{-20pt}
\end{deluxetable}

\begin{figure}[t]
	\centering
	\includegraphics[width=0.95\linewidth]{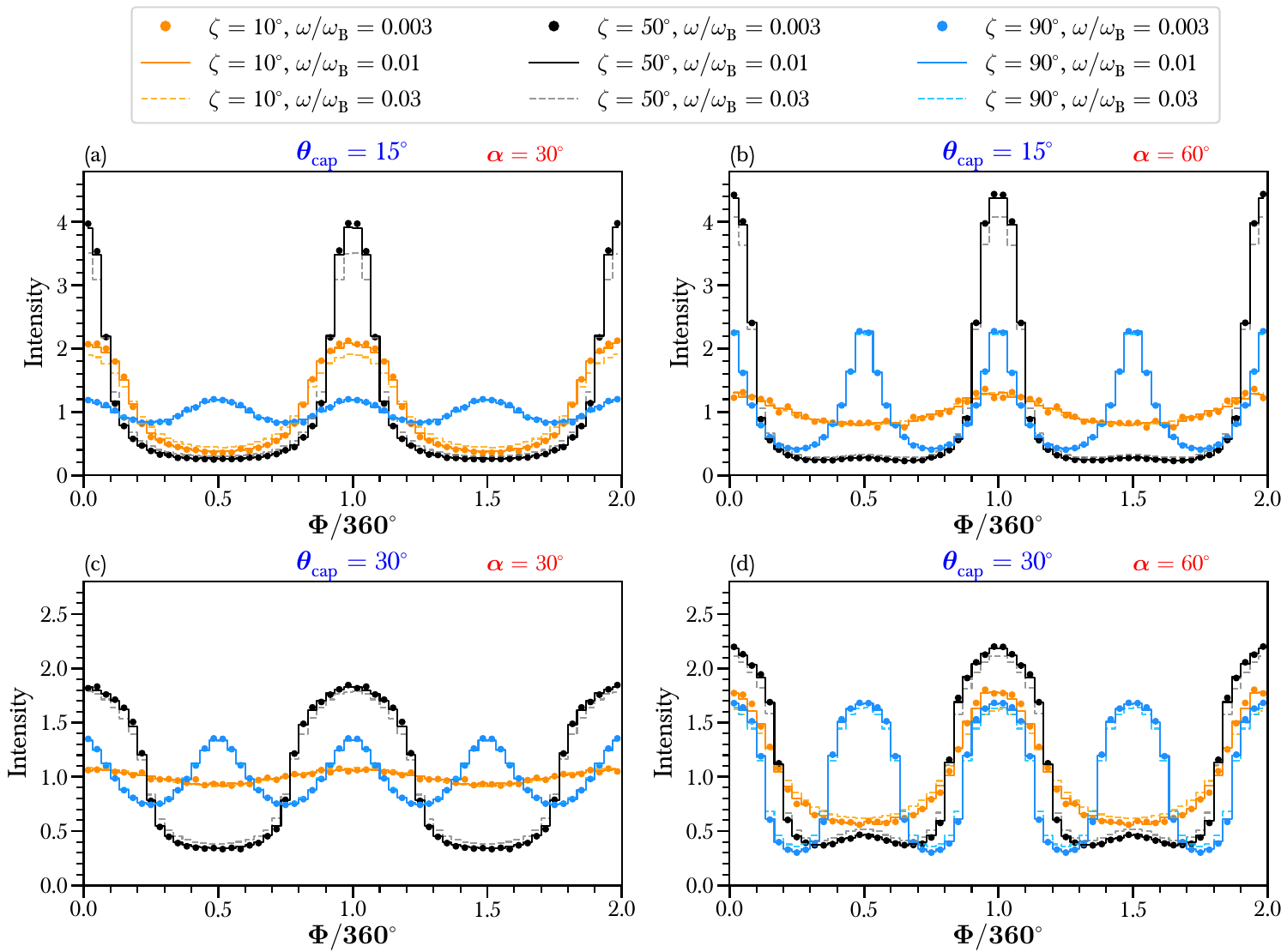}
	\caption{Simulated intensity pulse profiles from the {\sl MAGTHOMSCATT} simulation displayed for two rotational cycles for three different frequency ratios: $\omega / \wcyc = 0.003$ (dots), $\omega / \wcyc = 0.01 $(solid lines), and $\omega / \wcyc = 0.03$ (dashed lines). Emissions are from two identical antipodal polar caps, extending from the respective magnetic poles to colatitudes of $\theta_{\rm cap} = 15^{\circ}$ (top row) and $\theta_{\rm cap} =  30^{\circ}$ (bottom row); all photons are propagated through a general relativistic magnetosphere. Two values of inclination angles are considered: $\alpha = 30^{\circ}$ (left column) and $\alpha = 60^{\circ}$ (right column). Each panel shows the pulse profiles for three selected viewing angles:  $\zeta = 10^{\circ}$ (orange), $\zeta =  50^{\circ}$ (black), and $\zeta = 90^{\circ}$ (blue).   }
	\label{fig:PP_comparison_I}
\end{figure}

%\newpage

\subsection{Light curve saturation in the magnetar domain}
 \label{subsec:APstar-magnetar}

With this new, efficient injection protocol in hand, it is important to understand what the computational limitations are in the most problematic frequency domain, namely $\omega\ll \wcyc$ pertaining to magnetars, and furthermore what is actually needed to effectively model observational data.  A clue is provided by the polarimetric data presented in Figure~\ref{fig:Slab-2d}, specifically that the \teq{\theta_z -\phi_z} angular distributions do not change substantially once $\omega /\wcyc$ drops below 0.1.  Yet these results apply to just specific surface locales, rather than the ensemble of regions contributing to neutron star signals. To address soft X-ray data from magnetars, it is necessary to consider energy-dependent pulse profiles for different stellar geometrical parameters.  The geometry information is usefully captured in the $\alpha$-$\zeta$ ``sky maps,'' with $\alpha$ and $\zeta$ respectively being the angles between the magnetic axis
and the observer's viewing direction relative to the spin axis, for pulsed emission as exhibited in Figures 4 and~5 of \citet{Hu-2022-ApJ}.  This can be concisely presented for the new AP* protocol for an array of different frequency ratios via a selective survey of light curves, temporal variations of intensity and polarization for a rotating neutron star that an observer might measure.

\begin{figure}[t]
	\centering
	\includegraphics[width=0.95\linewidth]{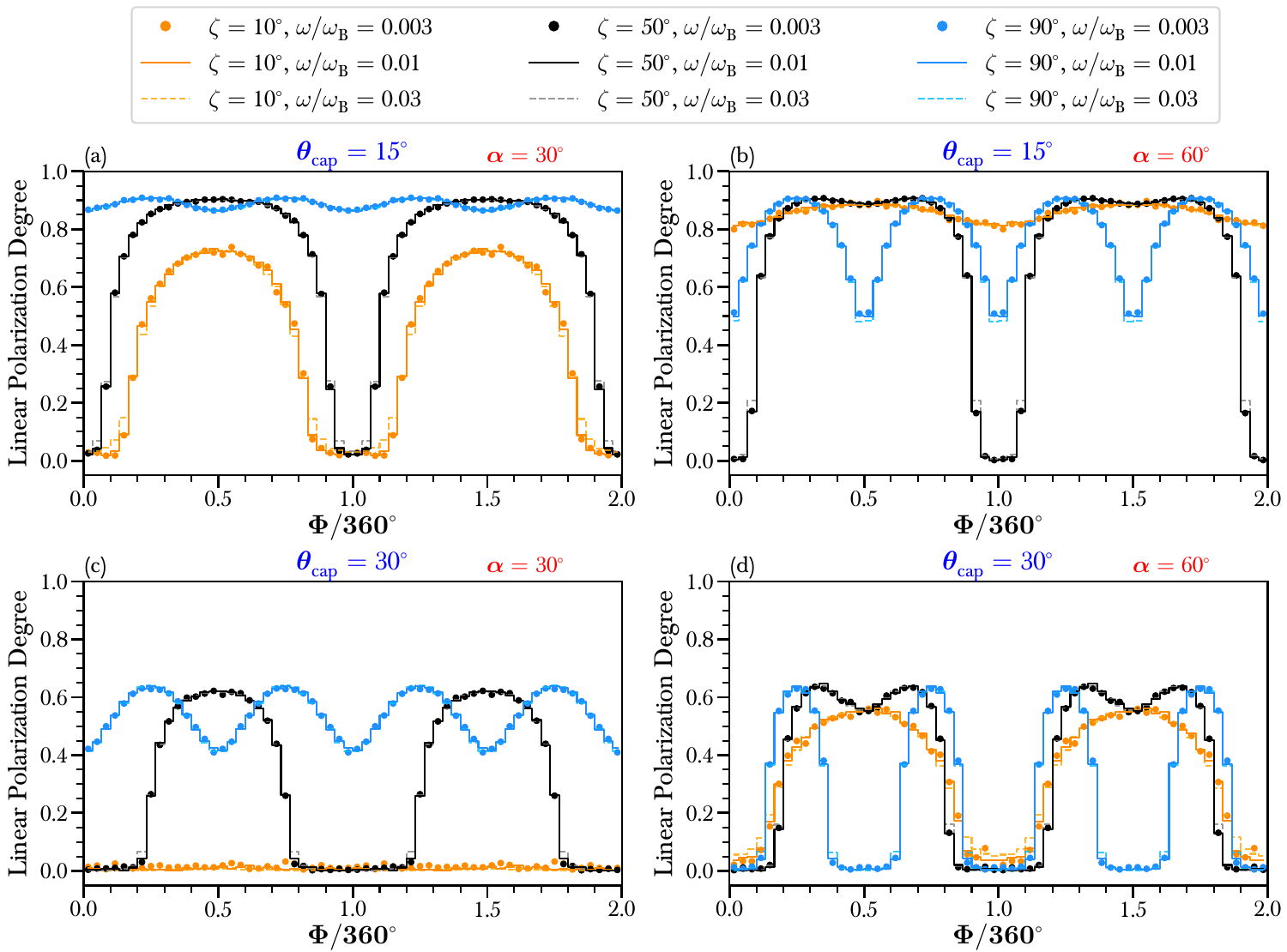}
	\caption{Same as Figure~\ref{fig:PP_comparison_I}, but for the linear polarization degree.}
	\label{fig:PP_comparision_Pi}
\end{figure}

To this end, using the AP* protocol described above, we performed simulations of X-ray emission from extended surface regions, propagating the emergent photons through a general relativistic magnetosphere as outlined in Section~4 of \citet{Hu-2022-ApJ}, for which plasma dispersion and vacuum birefringence effects on polarization during this propagation were omitted. Specifically, we modeled the emitting hot spots as two identical antipodal polar caps extending from their respective magnetic poles to colatitudes of $\theta_{\rm cap}$.  The mass and radius of the star were fixed at $M = 1.44$~$M_{\odot}$ and $R = 10$~km, respectively. Three different sub-cyclotronic frequency ratios were considered: $\omega / \wcyc = 0.03$, $\omega / \wcyc = 0.01$, and $\omega / \wcyc = 0.003$. For the case where $\theta_{\rm cap} = 15^{\circ}$, the respective numbers of photons recorded at infinity were $\mathcal{N}_{\rm rec} = 10^8$, $10^7$, and $3 \times 10^6$ from each cap. We decreased the number of recorded photons when decreasing $\omega / \wcyc$ in order to accommodate for the increase in run time. Particularly, the respective run times per $10^6$ photons for the three above frequency ratios, using the same computing setup, were 0.49 hours, 4.4 hours, and 29.5 hours.

The intensity and linear polarization degree pulse profiles for two selected inclination angles, $\alpha = 30^{\circ}$ (left panel) and $\alpha = 60^{\circ}$ (right panel), for $\theta_{\rm cap} = 15^{\circ}$ are shown in the top row of Figures~\ref{fig:PP_comparison_I} and \ref{fig:PP_comparision_Pi}, respectively. We also performed simulations for two annuli that span from colatitudes of $15^{\circ}$ to $30^{\circ}$ in the two hemispheres of the NS, using the same statistics as listed above. The results obtained with these annuli were then combined with those corresponding to $\theta_{\rm cap} = 15^{\circ}$ with proper scaling factors to the emission areas, yielding results for $\theta_{\rm cap} = 30^{\circ}$ polar caps. The latter are presented in the bottom row of Figures~\ref{fig:PP_comparison_I}-\ref{fig:PP_comparision_Pi}. Comparing the results obtained with $\omega / \wcyc = 0.03$ (dashed lines) and the other two frequency ratios, we can observe a relative difference of $~10{\%}$ in a few instances. Notably, the intensity pulse profiles for $\omega / \wcyc = 0.01$ (solid lines) and $\omega / \wcyc = 0.003$ (dots) are always almost identical, with relative differences of only $\sim 2\%$. { Similarly small differences are evinced in the polarization degree profiles for the three frequency ratios in Figure~\ref{fig:PP_comparision_Pi}.}  Such small differences are essentially unresolvable using extant observational pulse profiles.  Using full $\alpha$-$\zeta$ ``sky maps'' (not depicted), we verified that the close agreement between the results obtained for these two frequency ratios remain true across all viewing angles $\zeta$, ranging from $0^{\circ}$ to $180^{\circ}$, and inclination angles $\alpha$, from $0^{\circ}$ to $90^{\circ}$.  Such insensitivity to $\omega / \wcyc$ is consistent with our expectations based on the mathematical form of the magnetic Thomson cross-section in this high-field domain.  Indeed, in the domain where $\omega / \wcyc \ll 1$, the emission characteristics are sensitive to the magnetic field strength only within the MSC. For $\omega / \wcyc = 0.01$, the size of MSC is sufficiently small that further decrease in $\omega / \wcyc$--that is, increase in the magnetic field strength--does not alter the emergent configuration. Therefore, for the purpose of modeling magnetar data, it is unnecessary to conduct numerically inefficient simulations at frequency ratios lower than $\omega /\wcyc = 0.01$.

{ 
As mentioned in the Introduction, an important element of polarization considerations for magnetars is the influence of the birefringence of the magnetized quantum vacuum (VB) on the magnetospheric propagation of light.  This effect is not included in the results displayed in Figure~\ref{fig:PP_comparision_Pi}.  Yet, the design of {\sl MAGTHOMSCATT} in tracking the complex electric field vectors of photons is well suited to the incorporation of VB influences outside the star. The treatment of magnetospheric VB is addressed in detail in Dinh Thi et al. (in prep.), wherein it is shown to significantly increase the net linear polarization degrees that observers at infinity would detect.  Such an enhancement of polarization is also apparent in prior studies of magnetospheric birefringence \citep[e.g.,][]{Heyl-2000-MNRAS,Adelsberg-2006-MNRAS,Fernandez-2011-ApJ, Taverna-2015-MNRAS}. The improvements in code efficiency addressed in this paper also apply to the upgraded simulation that includes birefringent photon propagation above the stellar surface: the polarized propagation algorithm introduces only very small additions to the code run time.}

%\newpage

\section{Conclusions}
 \label{sec:conclusions}
In this paper, we present recent developments of our Monte Carlo simulation, {\sl MAGTHOMSCATT}, which models polarized radiative transfer in neutron star atmospheres using a complex electric field vector formalism. The simulation exhibits the Markovian nature of diffusion, where the anisotropy and polarization distributions at convergence are independent of the initial conditions and remain invariant with an increase in the number of scatterings as \teq{\tau_{\rm eff}} becomes larger.  By examining spatial distributions of photons within an atmospheric slab, we highlighted that exit tracks from the top and bottom of the simulated slab are inextricably linked in a way that makes it difficult to statistically select emergence from the upper atmosphere based on number of scatterings.
The emergent signatures of radiation were analyzed across different frequency domains and magnetic field orientations. These characteristics are consistent with the magnetic Thomson scattering cross section physics.
In the limit of non-magnetic field domain, $\omega / \wcyc \gg 1$, we compared our results for the distributions of intensity and linear polarization degree with those obtained by numerically solving  the non-magnetic radiative transfer integro-differential equation in \cite{Chandrasekhar-1960-book,Sunyaev-1985-AandA}. The excellent agreement serves as validation for the robustness of our simulation, complementing the validations presented in \citet{Barchas-2021-mnras}. Additionally, analytical fits for the non-magnetic profiles for intensity \teq{I} and Stokes \teq{Q} were provided: these have good precision and could have potential applications in numerical simulations and data interpretation, such as for modeling millisecond pulsars.
  
A particular emphasis was placed on improving efficiency in the domain where $\omega / \wcyc \ll 1$, which is pertinent to soft X-ray emission from magnetar surfaces. In this context, the disparity of the cross sections for different polarization modes and propagation directions necessitates higher optical depths for convergence, hence making the code inefficient. To address this, we proposed a protocol named AP*, which combines the AP injection protocol with a prescription for the minimum number of scatterings. The AP* protocol was shown to achieve convergence at lower optical depths, thereby accelerating the code by a factor of about \teq{3-8}, compared to the AP counterpart. 
Using the new efficient injection protocol AP*, we simulated soft X-ray emission from extended regions on neutron star surfaces for various sub-cyclotronic frequencies, considering the propagation of photons through a general relativistic magnetosphere. We found that the properties of polarization and intensity for $\omega / \wcyc \ll 1$ are well characterized by $\omega / \wcyc =0.01$ results. At this frequency ratio, the magnetic scattering cone $\theta_{\rm MSC} \lesssim \omega /\wcyc$ is sufficiently small that a further decrease in $\omega / \wcyc$  has a negligible impact on the detailed shape of pulse profiles for intensity and polarization. As a result, this circumvents the need to address slower runs at lower frequency ratios $\omega/\wcyc$. 
The approach outlined in Section~\ref{sec:new_protocol} will be applied in Dinh Thi et al. (in prep.), where we will compare our simulations with observational data from a bright magnetar 1RXS~J1708-4009. 

\begin{acknowledgements}

M.G.B. thanks NASA for generous support under awards 80NSSC22K0777, 80NSSC24K0589 and 80NSSC25K7257.  
This work was supported in part by the Big-Data Private-Cloud Research Cyberinfrastructure MRI-award funded by NSF under grant CNS-1338099 and by Rice University's Center for Research Computing (CRC).

\end{acknowledgements}


\vspace{5mm}
%\facilities{HST(STIS), Swift(XRT and UVOT), AAVSO, CTIO:1.3m,


%\software{astropy \citep{2013A&A...558A..33A,2018AJ....156..123A},  
  %        Cloudy \citep{2013RMxAA..49..137F}, 
     %     Source Extractor \citep{1996A&AS..117..393B}
    %      }


%% figure* and Table counter will not reset.

%\appendix


\bibliography{ref}{}
\bibliographystyle{aasjournal}



\end{document}